\documentclass[letterpaper]{article}
\usepackage{graphicx}
\newif\ifAMStwofonts

\def\degr{\hbox{$^\circ$}}
\def\arcmin{\hbox{$^\prime$}}
\def\arcsec{\hbox{$^{\prime\prime}$}}
\def\utw{\smash{\rlap{\lower5pt\hbox{$\sim$}}}}
\def\udtw{\smash{\rlap{\lower6pt\hbox{$\approx$}}}}

\begin{document}

\title[Absolute Parameters of Young Stars]{Absolute Parameters of Young Stars: GG Lup and $\mu^1$ Sco}

\author[E. Budding et al.]{E. Budding$^{1,2,3,4} $, R. Butland$^{5}$, M. Blackford$^{6}$  
\vspace{2mm} \\
$^{1}$ Department of Physics \& Astronomy, University of Canterbury;\\	
$^{2}$ Carter Observatory; 
$^{3}$ SCPS, Victoria University of Wellington, New Zealand; and \\ 
$^{4}$ Physics Department, University of Canakkale, TR 17020, Turkey.\\
$^{5}$ Visiting Astronomer, Dept. Physics \& Astronomy, University of Canterbury, New Zealand.\\
$^{6}$ Variable Stars South, RASNZ. \\
}
 
\maketitle

\begin{abstract}

New high-resolution spectroscopy and $BVR$ photometry, together with literature data,
on the Gould's Belt close binary systems GG Lup and $\mu^1$ Sco are presented and analysed. 

In the case of GG Lup, light and radial velocity curve fittings confirm a near-Main-Sequence 
picture of a pair of close stars. 
Absolute parameters are found, to within a few percent, thus: 
$M_1$ =  4.16$\pm$0.12, $M_2$ = 2.64$\pm$0.12,
$R_{1}$ = 2.42$\pm$0.05, $R_2$ =  1.79$\pm$0.04, ($\odot$);
$T_{1}$ $\sim$13000,  $T_2$ $\sim$10600 (K); 
photometric distance  $\sim$ 160 (pc). 
The high eccentricity and relatively short period (105 y) of apsidal
revolution may be related to an apparent
`slow B-type pulsator' (SPB) oscillation.  Disturbances of the
outer envelope of at least one of the components
then compromise comparisons to standard evolutionary models,
at least regarding the age of the system.
A rate of apsidal advance is derived, which allows 
a check on the mean internal structure constant 
$\overline{k_2} = 0.0058 \pm 0.0004$.  This is in
agreement with values recently derived for young
stars of solar composition and mass $\sim$3${\odot}$.
  
  For $\mu^1$ Sco, we agree with previous authors that the secondary
  component is considerably oversized for its mass, implying binary (interactive) stellar evolution, probably of the `Case A' type.  The primary appears relatively little affected by this evolution, however.
 Its parameters show consistency with a star of its derived mass 
 at age about 13 My, consistent with the
 star's membership of the Sco-Cen OB2 Association.
 The absolute parameters are as follows:
$M_1$ =  8.3$\pm$1.0, $M_2$ = 4.6$\pm$1.0,
$R_{1}$ = 3.9$\pm$0.3, $R_2$ =  4.6$\pm$0.4, ($\odot$);
$T_{1}$ $\sim$24000,  $T_2$ $\sim$17000 (K); 
photometric distance  $\sim$ 135 (pc). 


\end{abstract}

\label{firstpage}


\section{Introduction}
Our work on young southern binary systems has been 
presented in several earlier papers concentrating 
on individual well-documented examples.  In the present paper
we consider two close binaries -- GG Lup and $\mu^1$ Sco --
that allow more compact summaries.
Further background  was given by Budding, (2008 -- hereafter `Paper
I').  A recent general review was given by Idaczyk et al.\ (2013).  Andersen's (1991)
review gave an interesting rationale for this kind of work.
 
We give more particulars on GG Lup next.
Photometric data (Section 2.1) are later combined with 
new spectroscopic material (Sections 2.2-5) to refine knowledge
of the absolute parameters of the components (Section 2.6). 
A similar treatment then follows for $\mu^1$ Sco in Section 3.
Section 4 summarizes the derived information on both binaries within the context of their galactic environment.
This arrangement follows similar lines to
that of previous papers of this programme.

Although these two young, short-period, early type systems are both
likely members of the Sco-Cen OB2 Association and of comparable age,
their physical characteristics are quite different in 
some respects. They do, however, both show indications of 
additional variability, beyond that associated with close binarity.
Such effects compromise their precise parametrization.
We aim to clarify the background on their observed properties.
At the same time, we challenge some of 
the claims of previous authors on modelling details.

\section{GG Lup}

GG Lup (= HD 135876, HIP 74950, HR 5687) is a relatively bright
($V$ $\sim$ 5.5-6.1, $B - V$ $\approx$ $-0.099$, 
$U - B$ $\approx$ $-0.46$, $V - I$ $\approx$ $-0.08$,
$V - J \approx -0.530$,
$V - H \approx -0.667$,
$V - K \approx -0.625$; Cutri et al., 2003), young B7V + B9V type
close binary system with period $P \approx 1.85$ d, having a known appreciable 
eccentricity $e \approx 0.15$.
Neubauer (1930) discovered it as a close spectroscopic pair.
Photometry showing eclipses later appeared from Smith (1966)
and Strohmeier (1967).  A thorough study was given by Andersen et al.\ (1993) 
who, noting the youth of the stars (age estimated at $\sim$20 My), 
found a relatively short period of just over 100 y for its apsidal advance.
This was confirmed by Wolf and Zejda (2005), although the coverage of the
O -- C  curve is still not great.

The sky location RA(2000) = 15h 18m 56s; dec(2000) = --40\degr 47\arcmin 18\arcsec, 
HIPPARCOS distance 162 $\pm$ 17 pc and proper
motions ($\mu_{\alpha} \cos \delta \approx -24$, $\mu_{\delta} \approx -22$ mas y$^{-1}$) make the system
a likely member of the Upper Centaurus Lupus (UCL) 
 concentration
(Blaauw, 1964) of the Sco-Cen OB2 Association,
within the Gould's Belt giant star formation region (Nitschelm,
2004). The system lies towards that part of the Belt
that is more separated from the Disk
($l$ = 330.9; $b$ = 14.0 deg). The relatively scant attention
given previously to this interesting object may be due to its
southerly declination ($\sim -40$ deg).  

Wolf  and Kern (1983) obtained {\em uvby}$\beta$
measures for GG Lup, noting a need for a revised ephemeris,
while Levato and Malaroda (1970), at La Plata Observatory, published a
fairly high projected rotational velocity of 155 km s$^{-1}$, presumably of the primary component.
Andersen et al.\ (1993) emphasized
the implications of their parameters for the system on the
relationship of stellar properties to galactic environment: an issue
that has significance to problems of stellar formation in general
and Gould's Belt in particular.

\subsection{Photometry and analysis}
 
We start with an examination of the HIPPARCOS light curve (ESA, 1997).
The HIPPARCOS data source gives the ephemeris
 Min I = 2448500.55 + 1.84962E,
which can be compared with that of Andersen et al.\ (1993)
 Min I = 2446136.7398 + 1.84960E.  Kreiner et al (2001) refine the ephemeris to 
  Min I = 2446136.7477 + 1.8495927E.
We show the HIPPARCOS light curve in Fig~1 together with our best-fitting model,
whose main parameters are listed in Table~1.
This modelling was done using the program {\sc fitter}, whose theoretical basis
was discussed by Budding \& Demircan (2007; Ch.\ 9). The program's operation
(recently upgraded to {\sc winfitter}) has been detailed by
Rhodes (2014), {\bf while Rhodes \& Budding (2014) presented a 
detailed discussion of the modelling and fitting procedure.}
 The approach uses a standard Marquardt-Levenberg type optimization
technique with a fitting function that derives from Kopal's (1959) linearization
of close binary proximity effects in terms of spherical harmonics.
The tidal and rotational distortions thus include the effects of finite
mass of the envelopes.  The reflection terms provide an
`albedo' factor (usually unity for radiative envelopes), 
for possible empirical modification of the 
luminous efficiency coefficients of Hosokawa (1958).
 
The mean epoch of the HIPPARCOS observations is given as HJD 2448349.0625, almost 
six months before the time of minimum given above, and the data was collected over
an approximately 3 year interval, so some uncertainty in the timing of the
relevant periastron longitude arises. A more representative average epoch  
for the HIPPARCOS data
would  be HJD 2448348.881; some 82 periods before the cited one.

\begin{figure}
\label{fig-1}
\includegraphics[height=8cm,angle=-90]{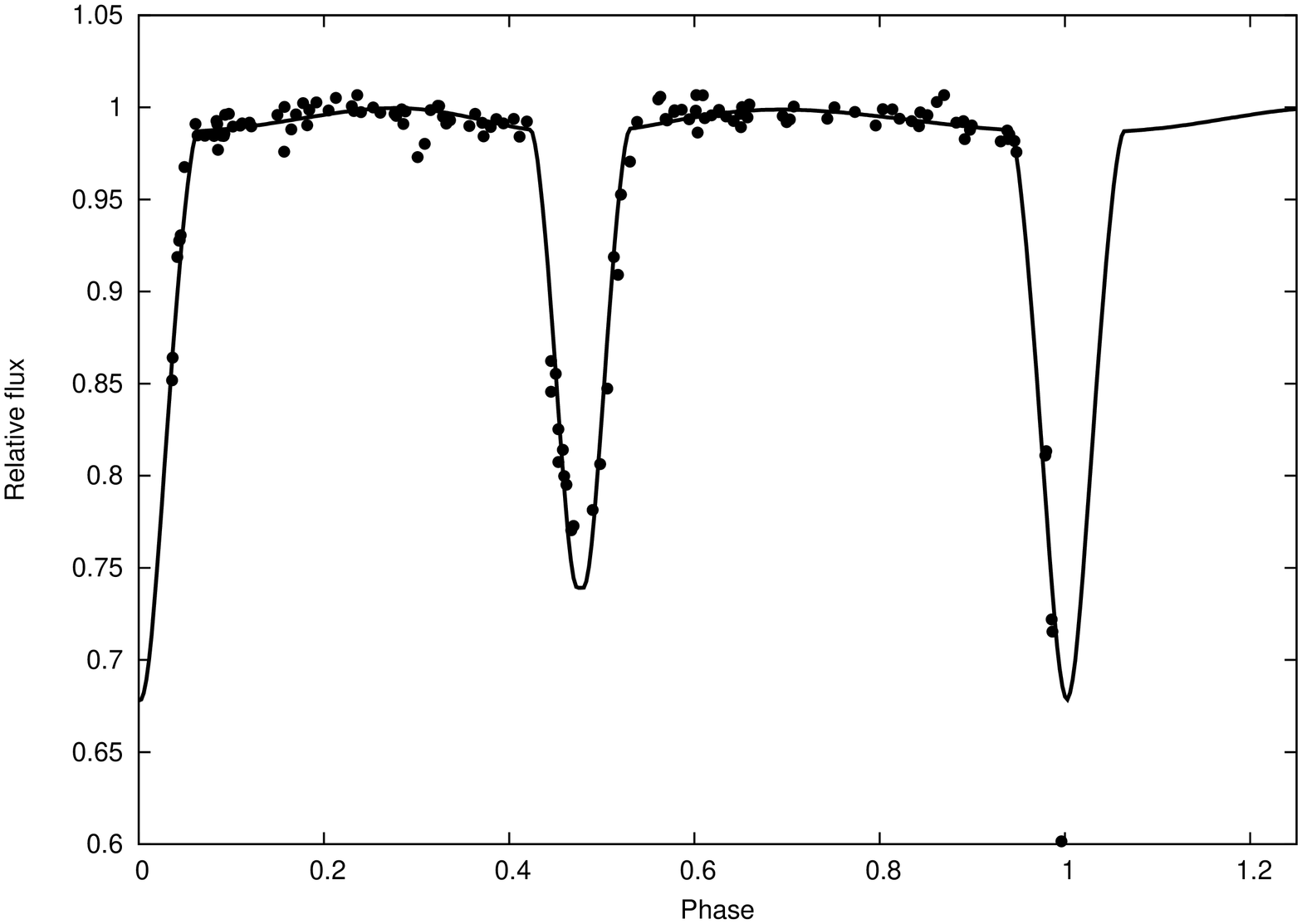}\\
\includegraphics[height=8cm,angle=-90]{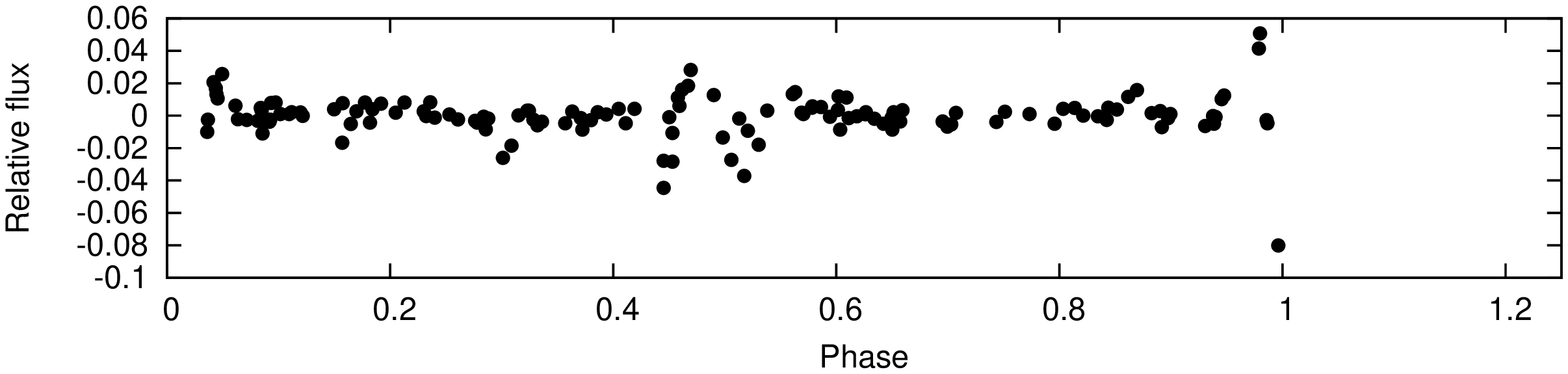}\\
\caption{HIPPARCOS V photometry of GG Lup: model fitting and residuals
(see Section 2 for details).}
\end{figure}

\begin{table}
\begin{center}
\caption{Curve fitting results for HIPPARCOS photometry of GG Lup.
Parameters for which no error estimate is given are adopted from
information separate to the fitting.
\label{tbl-1}} 
\begin{tabular}{lcc}
   \hline  \\
\multicolumn{1}{c}{Parameter}  & \multicolumn{1}{c}{Value} & 
\multicolumn{1}{c}{Error}\\
\hline \\
$T_h$ (K) & 12000 & \\
$T_c$ (K) & 10000 & \\
$M_2/M_1$ & 0.62 & \\
$L_1$ & 0.79 & 0.01 \\
$L_2$ & 0.21 & 0.01 \\
$r_1$ (mean) & 0.204 & 0.003\\
$r_2$ (mean) & 0.141 & 0.004 \\
$i$ (deg) & 88.3 & 1.4 \\
$e$  & 0.15 &\\
$\omega$ & 108.1 & 0.4\\
$u_1$ & 0.35 &\\
$u_2$ & 0.41 &\\
$\Delta\phi_0$(deg) & --2.07 & \\
$\chi^2/\nu$ & 1.05 & \\
$\Delta l $& 0.007  &\\
\hline
\end{tabular}
\end{center}
\end{table}

Turning to the $u v b y$ data of Clausen et al.\ (1993)
as cited by Andersen et al.\ (1993), we provide the results of 
curve-fitting experiments in Table~2, with corresponding diagrams in Figs~2 and 3.  Table~3 provides the adopted reference mean magnitudes, in various filter regions,
 using fittings for the relative luminosities as shown
 in Table~2.  Andersen et al.\ (1993) gave a value
 of 0.006 mag as a representative error for their 
 photometry.  With the added uncertainties of the fittings
 and the non-uniformity of the magnitude scale this translates to $\sim$0.01 mag for the brighter component and
 $0.05$ mag for the fainter star.

\begin{figure}
\label{fig-2}
\noindent\includegraphics[height=6cm,angle=0]{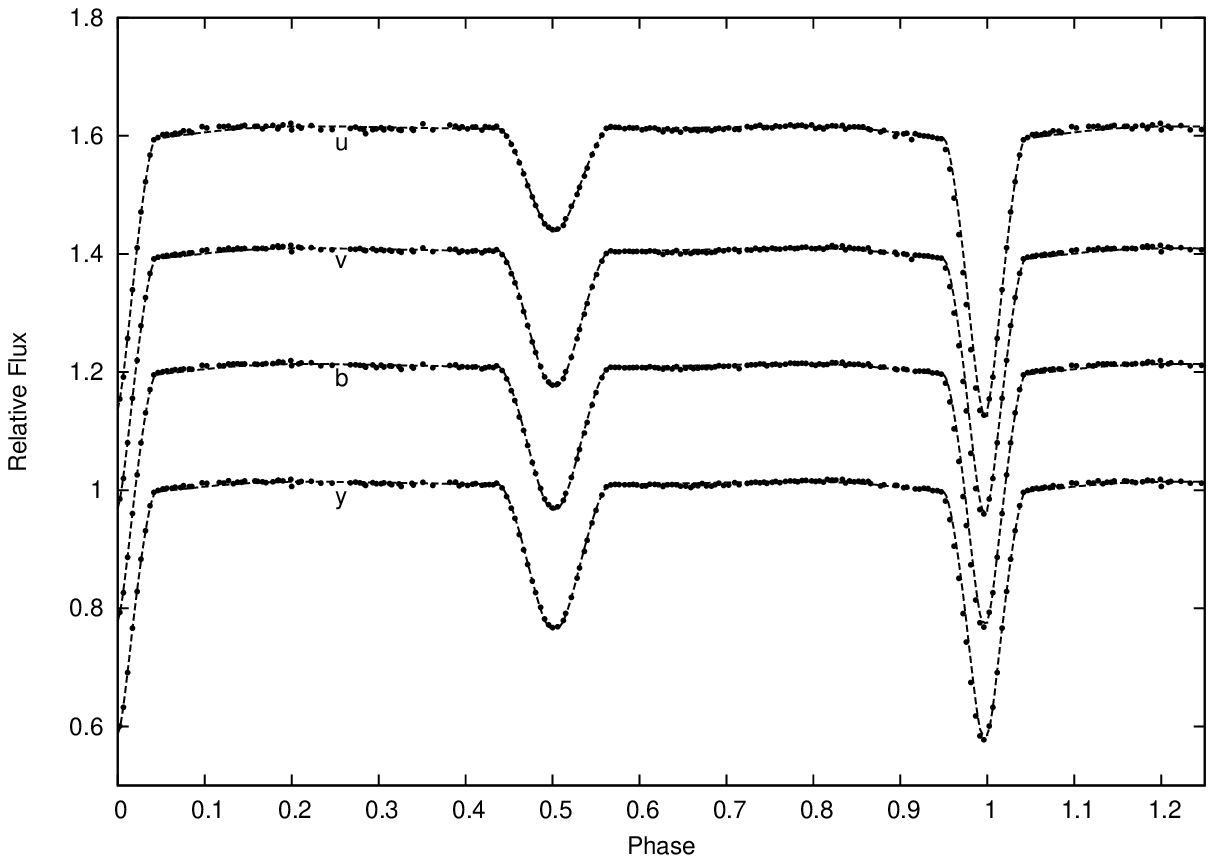}\\
\caption{Clausen et al's (1993) $u v b y$ photometry of GG Lup and model fitting.  Residuals are shown in Fig 5.
}
\end{figure}

{\small
\begin{table}
\begin{center}
\caption{Curve fitting parameters for Clausen et al's (1993) $u v b y$ photometry of GG Lup.  Error estimates are shown by the bracketed
numbers, that relate to the final digit of the parameter values.
Otherwise, parameter valued are adopted.
\label{tbl-2}} 

\begin{tabular}{lccccc}
   \hline  \\
\multicolumn{1}{c}{Parameter}  & \multicolumn{5}{c}{Value} \\
\multicolumn{1}{c}{} & \multicolumn{1}{c}{mean} & \multicolumn{1}{c}{$u$} & \multicolumn{1}{c}{$v$}  & \multicolumn{1}{c}{$b$} & \multicolumn{1}{c}{$y$} \\
\hline \\
$T_h$ (K) & 12000 & & & & \\
$T_c$ (K) & 10000 & & & & \\
$M_2/M_1$ & 0.62 & & & & \\
$L_1$ & & 0.841(7) & 0.776(6) & 0.755(6) & 0.747(6)   \\
$L_2$ & & 0.159(5) & 0.224(4) & 0.245(4)  & 0.253(4)   \\
$r_1$  & 0.201(1) & & & & \\
$r_2$  & 0.149(1) & & & & \\
$i$ (deg) & 87.5(4)  & &  & & \\
$e$  & 0.154 & & & &\\
$\omega$ (deg) & 86.3(1) & & & &\\
$u_1$ & &0.43 &0.43 &0.40 & 0.35 \\
$u_2$ & &0.46 &0.50 &0.47 & 0.41\\
$\Delta\phi_0$(deg) & 1.63 & & & &  \\
$\chi^2/\nu$ & 1.13 & & & &  \\
$\Delta l $& 0.0025 & & & & \\
\hline
\end{tabular}
\end{center}
\end{table}
}

\begin{figure}
\label{fig-3}
\noindent\includegraphics[height=6cm,angle=0]{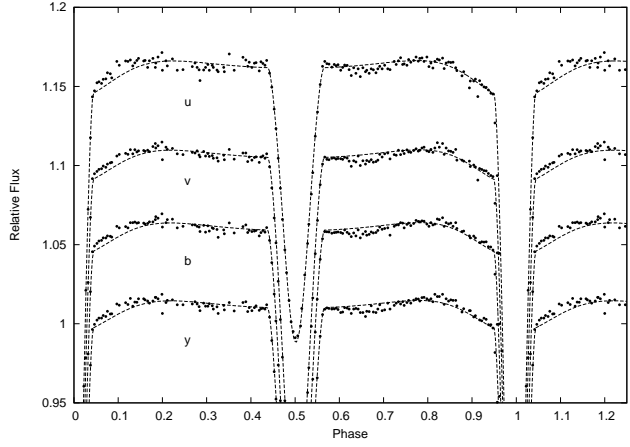}\\
\caption{$u v b y$ photometry with enlarged ordinate scale,
showing the oscillatory behaviour mentioned by Andersen et al.\ (1993).}
\end{figure}

\begin{table}
\begin{center}
\caption{Magnitudes of stars in the GG Lup  system.
Errors are discussed in Section (2.1).
\label{tbl-3}} 
\begin{tabular}{crrrr} 
\multicolumn{1}{c}{} & \multicolumn{1}{c}{$u$} &\multicolumn{1}{c}{$v$}
& \multicolumn{1}{c}{$B$} & \multicolumn{1}{c}{$b$} \\ 
$m_{\rm sys}$ & 6.194  &  5.612 &  5.549 & 5.547 \\
$m_1$         & 6.382  &  5.887 &  5.733 & 5.852 \\
$m_2$         & 8.19   &  7.24  &  7.09  & 7.07  \\
 &  & & & \\
\multicolumn{1}{c}{} & \multicolumn{1}{c}{$y$} &\multicolumn{1}{c}{$V_H$}
& \multicolumn{1}{c}{$V$} & \multicolumn{1}{c}{$R$} \\
$m_{\rm sys}$ & 5.589 &  5.577 &  5.558 & 5.613 \\
$m_1$         & 5.906  & 5.847  & 5.853  & 5.911 \\
$m_2$       &   7.08   & 7.22   & 7.12   & 7.16  \\
\end{tabular}
\end{center}
\end{table}

\subsubsection{Oscillatory effects}
Andersen et al.\ (1993) referred to systematic out-of-eclipse variations, visible in the enlarged scale of Fig~3,
as residual inherent variation of the source that could not be removed by eclipsing binary model parameter adjustments. They mentioned forced oscillations, tidal lag or non-aligned rotation axes as possible causes.
They did not pursue such effects, however, 
concentrating their study on the determination of basic stellar absolute parameters. 
Tidal lag or non-aligned rotation axes can be seen as somewhat particular examples 
of general dynamical effects for the fluid masses making up this 
close pair (probably not very effective, taken individually, in accounting for 
the scale and apparent periodicity of the effects).

We may thus consider the apparent three-to-one resonance between the oscillatory and orbital photometric variations 
within the context of forced effects associated with the unusually high eccentricity of the system 
and some inherent overstability of at least one of the component stars.
Lampens (2006), noting the special relevance of young star associations in this context, 
considered the sequence of interactions involving orbital motion, rotation and pulsation as an 
interesting question, still open to detailed study.
An interesting review on this subject was given by Harmanec et al.\ (1997), as an introduction to their concentrated work on the comparable, but much longer period eclipsing binary system V436 Per. Chapellier et al.'s (1995) study of the early type eccentric
eclipsing binary EN Lac is also noteworthy in this context. 
Orbital circularization in early type close binaries is expected to be a relatively fast process (Zahn, 1977), i.e.\ $\tau_{\rm circ} < 10^6$ y, in general.  This process would probably involve dissipative mechanisms related to $g$-mode oscillations (Henrard, 1982; Alexander, 1987,88; Witte \& Savonije, 1999).
 
If there already exists a $\kappa$-mechanism capable of generating inherent non-radial pulsations, 
particular arrangements of the eccentricity that would otherwise produce the same or similar modes may become trapped, 
since the dissipation affecting such modes would, in this case, be accounted for by the $\kappa$-mechanism.  
Work otherwise drawn from the reserve of orbital energy to cover dissipation of the excited fluid motions 
now comes from another source and the corresponding secular decline of eccentricity halted. 
Such motions are, from an idealized classical mechanics point of view, 
as capable of putting energy
back into the orbit as being stimulated by that. But the configurations that would be more likely observed
would be the quasi-static ones, where the oscillation driving just matches secular dissipation.
Relative minima of the net disturbing potential would then be expected, out of which the orbit would 
be gradually perturbed by evolutionary changes of the
$\kappa$-mechanism.  In the case of SPB oscillations, these can be expected over the 
star's Main Sequence lifetime.  But Lampens (2006) notes, pointedly, that six of the seven 
short period SPBs found in close binary systems  by De Cat \& Aerts (2002) were in eccentric orbits.

\begin{figure}
\label{fig-4}
\noindent\includegraphics[height=9cm,angle=-90]{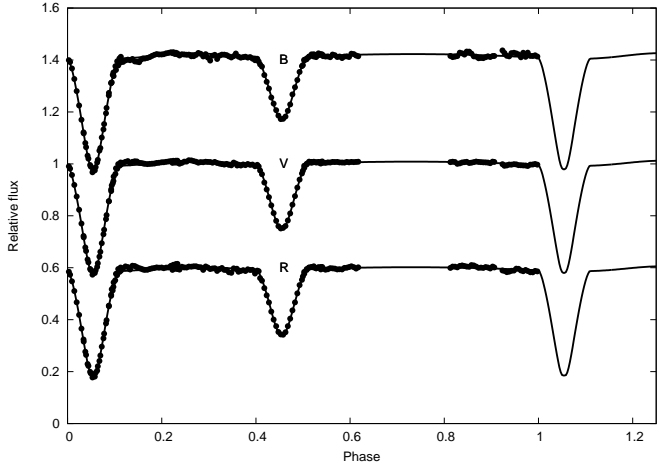}\\
\caption{DSLR photometry of GG Lup and model fitting.
Residuals are shown in Fig 6.}
\end{figure}

The criteria discussed by De Cat \& Aerts (2002) for establishing 
the SPB nature of a star usually involve at least two separate
photometric samplings together with spectroscopic evidence, and 
one of the photometric sources they refer to is HIPPARCOS.  In our present case, however, the amplitude of the photometric oscillation appears relatively small at $\sim$0.005 mag, and this
is appreciably below the scatter of residuals ($\sim$0.01 mag) in the HIPPARCOS light curve.  We were therefore interested to
check the photometric indications shown by Andersen et al.\ (1993) by an alternative technique and the application of 
a modern digital single-lens reflex (DSLR), high pixel count camera 
to this (Blackford \& Schrader, 2011) was opportune.

\subsubsection{New photometry}
Times of minima and {\em BVR} photometry of GG Lup and other stars have been conducted under the Southern Eclipsing Binaries Programme of the Variable Stars South (VSS) section of the Royal Astronomical Society of New Zealand (RASNZ). The GG Lup data were gathered using an unmodified Canon 450D (Digital Rebel XSi) and Nikkor 180 mm lens operated at f4. An unmodified Canon 600D (Digital Rebel T3i) and Canon 200 mm lens operated at f4 were used to collect similar data more recently.

\begin{figure}
\label{fig-5}
\noindent\includegraphics[height=6cm,angle=0]{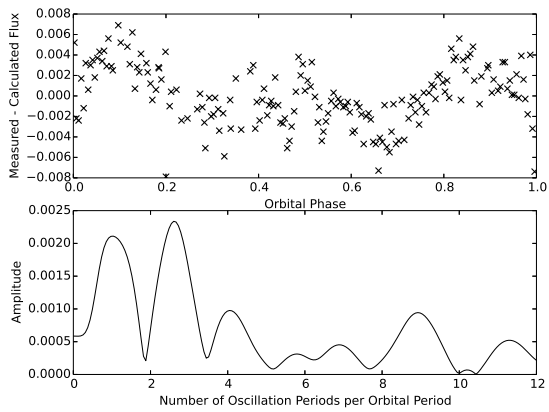}\\
\caption{Andersen et al.'s (1993) $y$ residuals,
with periodicity analysis shown in the lower panel.}
\end{figure}

\begin{figure}
\label{fig-6}
\noindent\includegraphics[height=6cm,angle=0]{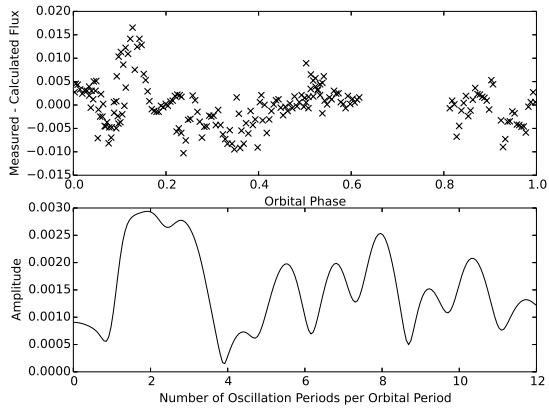}\\
\caption{Our $V$ photometry residuals treated in the
same way as for Fig 5.  The pulse around phase $\sim$0.15 may suggest
a `heartbeat' effect (Hambleton et al., 2013).}
\end{figure}

The Bayer filter array employed in DSLR cameras allows simultaneous recording of three colour ({\em rgb}) images, from which separate single colour images can be extracted and calibrated. From these images the instrumental magnitudes are determined by differential aperture photometry using an ensemble of comparison stars (6 for GG Lup). Atmospheric extinction correction was performed and instrumental magnitudes transformed to the standard (Johnson $BVR$) system. Our GG Lup light curves are shown in Fig~4, where we also show optimal fits to the reduced $B$, $V$ and $R$ datasets.
 
The $V$ residuals were subject to a frequency analysis using 
purpose-built software carrying out a Lomb-Scargle  and the results displayed in Figures 5 and 6.  Apart from any indications of additional oscillatory behaviour that the eye can detect in the upper panels
of these diagrams, the lower parts 
confirms the quasi-resonant $P/3$ character in the residual
variations of Clausen et al's (1993) data (Fig~5), and is 
suggestive, though not strongly corroborative,
of such behaviour in our own more recent data (Fig~6).

The effects of orbital eccentricity are clear in the light curves shown in Figs~1-4. Those produced around 1985 (Clausen et al., 1993) or 1991 (HIPPARCOS)
show the secondary minimum occurring about half-way between the primary minima, though they are distinctly wider. The line of apses must then be close to the line of sight. The recent DSLR data shown in Fig~4, on the other hand, show the separation between primary and the following secondary minimum (of very similar widths) to be significantly less than that between secondary and following primary.  This implies the line of apses now lying towards  perpendicular to the line of sight.   The elliptic orbit of the binary must then have rotated by about one quarter of a complete cycle in the years between the earlier and more recent light curves.  The details of the longitude of periastron variation are given in 
Table~4.  The derived mean rate of apsidal advance should not have the same precision as that corresponding to many individual timings of minima, of course; even so, the
fittings of a complete light curve, when gathered over a relatively short time, allows the periastron longitude to be accurately determined.  Our best estimate for the rate of apsidal advance (one revolution in 106$\pm$2 y) is not so far from Andersen et al.'s (1993) value of 101$\pm$4 y, which was matched by Wolf \& Zejda (2005), who found 101.6$\pm$1 y, essentially defined by a small number of photoelectric minimum timings over 
a $\sim$10 y interval. 

The implications of the derived rate of apsidal advance on the internal structure of the component stars may be checked following well-known procedures
(cf.\ e.g.\ Budding, 1974).  The observed rate of apsidal advance has first to be corrected for a relativistic effect which,
using the parameters of Table~7, amounts to some 8.16\% of the ratio 
of the orbital to proper apsidal periods $P/U$ = 4.525$\times$10$^{-5}$.   Using also the geometrical 
parameters of Table~2 and the rotation speeds discussed in section 2.5,
together with the formulae in Section 4 of Budding (1974), we can derive
a value for the mean structural constant (for the two stars) of $\overline{k_2}$ = 0.0058 $\pm$ 0.0004,
which compares favourably with the theoretical value of 0.0056
calculated by \.{I}nlek and Budding (2012) for a young 
star of 3 M${\odot}$ and near solar composition,
cf.\ also Claret and Gimenez (2010), who give further
background on this subject.  The independent calculations of 
\.{I}nlek and Budding (2012) for $\overline{k_2}$  are generally within $\sim$10\% of those of Claret and Gimenez (2010) for
comparable stellar models.  Although our value for  $\overline{k_2}$ is within Claret and Gimenez'  (2010) error limit  for their theoretical value, it is outside Andersen et al's (1993) error limit for their own empirical 
determination.

\begin{table}
\begin{center}
\caption{Apsidal advance from the 
light curve fittings (Clausen et al., 1993; HIPPARCOS and present study).
\label{tbl-4}} 
\begin{tabular}{lc}
\multicolumn{1}{c}{Epoch}  & \multicolumn{1}{c}{$\omega$}\\
\multicolumn{1}{c}{JD 2440000 +}  & \multicolumn{1}{c}{($\pm$)} \\
\hline 
6136.7398 & 86.3(1) \\
8348.881 & 108.1(4)  \\
16069.189 & 178.7(1)  \\
\hline
\end{tabular}
\end{center}
\end{table}

\subsection{Spectroscopy} 

As in previous similar work (Paper I, Budding et al., 2013)
our spectroscopic data were gathered using the
High Efficiency and Resolution Canterbury University Large
\'{E}chelle Spectrograph (HERCULES) of the Department of
Physics and Astronomy, University of Canterbury, New
Zealand (Hearnshaw et al., 2002).
 This was used with the 1m McLellan telescope at
the Mt John University Observatory (MJUO) near Lake
Tekapo ($\sim$ 43{\degr}59{\arcmin}S, 174{\degr}27{\arcmin}E).  Some 35 spectra were taken over the period May 14-19, 2006,
in fairly clear weather (occasional clouds). The 50 $\mu$m optical
fibre was mostly used, enabling a resolution of $\sim$70000.
The CCD (SITe) camera  was in  position 2 (Skuljan, 2004).
Typical exposure times were about 500 s. 
The data have been reduced using the software {\sc hrsp} version 5
(Skuljan, 2012) that produces output conveniently in {\sc fits} formatted files.
 
 About 40 clear orders could then be displayed
and studied using the program library of {\sc iraf}. 
 One of the {\sc iraf} subroutines ({\sc splot}), for example,
allows image statistics to be checked.  We could determine in this way
a signal to noise ratio (S/N) for continuum pixel regions (away from
flaws or telluric effects) to be usually of order 100 (with a 
$\sim$10\% deterioration from orders 85 to 121: the orders 
examined in this study).
Lines that could be measured well (see Table~5) have a depth
of typically 10\% of the continuum and a width at base of up to
 $\sim$7 {\AA} or $\sim$250 pixels. The positioning of a
well-formed symmetrical line can thus be expected to be
typically achieved to within a few pixels or (equivalently)
several km s$^{-1}$.  This agrees with the scatter
in derived radial velocities (rvs) given below.

{\sc iraf} also allows the equivalent width (ew) of features
to be directly read from {\sc splot}'s profile-fitting tool (`k').
We could check, in this way, the spectral type assignations
accepted in the literature (although often carried out with 
different techniques, dispersions and using different features
than those available to us).  Measured ratios of the 
H$_{\beta}$/He~I 6678 ews were typically $\sim$30.
This compares favourably with a mean spectral type of B8,
using the classical guides (e.g.\ Jaschek \& Jascheck, 1987)
in combination with the synthetic spectra of Gummersbach and Kaufer 
(2014). It follows that the secondary must be lower
in spectral type than the primary, from comparing the relative
He~1 6678 ews of primary and secondary $\sim$1.8, given that the continuum luminosity ratio is $\sim$3 in this part of the spectrum,
although the implied difference cannot be greater than two spectral classes.

The reduced high-dispersion data were 
manipulated as numerical text files, so that
 individual lines could be fitted with standard `dish-shaped' profiles
 (convolved with a Gaussian broadening) and the radial velocity determined from the profile centroids.  Useful results were obtained from the selection of
 lines listed in Table~5, particularly the He I lines.
Identifiable spectral lines for GG Lup are similar to those listed in Paper I. Certain secondary lines that are
normally too weak to measure reliably became 
relatively strong during the primary minimum phases
(e.g.\ Fe~II  5041.074).
Our profile fitting software package ({\sc spectrum}) was used to
derive line-related parameters for the component stars.  
  {\sc spectrum}, a {\sc python} module, uses functions from {\sc scipy}
 to display spectral lines and fit various user-defined functions
 to the measured data.    In particular, from {\sc scipy.optimize}
the `curve{\_}fit' function,  applying non-linear least squares optimization  principles, is used for the fitting as it is tolerant of noisy data.

\subsection{Radial velocities}

\begin{table}
\begin{center}
\caption{Spectral lines of GG Lup which were used for radial velocity determination.
\label{tbl-5}}
\begin{tabular}{ccll}
  & & &  \\
\multicolumn{1}{c}{Species}  & \multicolumn{1}{c}{Order no.} &
\multicolumn{1}{l}{Wavelength} & \multicolumn{1}{l}{Comment}  \\
He I & 85        &  6678.149  &   strong, p, usable s       \\
Si II & 89       &  6371.359  &  noisy p, unreliable s  \\
Si II & 95       &  5978.97  & weak p, unreliable s \\
He I &  97       &  5875.65  &  strong p, usable s\\
Fe II & 107      & 5316.693  & moderate p, weak s \\
Mg I  & 110      &  5183.604  & used for s, but weak   \\
Fe II & 110      &  5169.030  &  weak p, very weak s   \\
Fe I & 113      &  5041.074  & moderate p, weak s    \\
He I & 121       &  4713.258  &  (blend) moderate p, weak s  \\
\hline \\
 
\end{tabular}
\end{center}
\end{table}

The mean wavelengths derived from profile fitting to 
the lines in Table~5 allow representative 
rvs of the stars to be found, using the Doppler displacement principle.
The full schedule of rv observations is given in Table~6.
  The listed dates and velocities and have been corrected
to heliocentric values as in Paper I (using
{\sc hrsp} and {\sc iraf} data reduction tools).  
The precision of line-positioning was
discussed above.  It depends essentially on the
line properties, rather than the available 
spectrographic resolution which is high for HERCULES
compared with previous studies.  
 So while the rvs are given to one significant
decimal place in Table~6, in reality, even the digit before
the decimal point should be regarded with caution. 

\begin{table}
\begin{center}
\caption{Radial velocity data for GG Lup.
\label{tbl-6}} 
   \begin{tabular}{cccc} 
\multicolumn{1}{c}{No}  & \multicolumn{1}{c}{Phase} &
\multicolumn{1}{c}{RV1} & \multicolumn{1}{c}{RV2} \\
\multicolumn{1}{c}{}  & \multicolumn{1}{c}{d} &
\multicolumn{1}{c}{km s${-1}$} & \multicolumn{1}{c}{km s$^{-1}$} \\
\hline \\
 1 & 0.0451    &    ---  &   28.1 \\
 2 & 0.1163 &	--64.3  &  123.3 \\ 
 3 & 0.1516 &	--103.7  &  169.8 \\
 4 & 0.1758 &	--118.5  &  192.5  \\
 5 & 0.1768 &	--124.8  &  198.8 \\
 6 & 0.2522 &	--143.4  &  239.3 \\
 7 & 0.2554 &	--142.4  &  236.5  \\
 8 & 0.2590 &	--143.4  &   ---  \\
 9 & 0.2847 &	--139.0  &  230.7   \\
 0 & 0.3066 &	--127.3  &  210.0   \\
11 & 0.3096 &	--127.6  &  202.0   \\
12 & 0.3243 &	--119.0  &  193.6   \\
13 & 0.3378 &	--112.6  &  181.8   \\
14 & 0.3402 &	--115.2  &  177.8  \\
15 & 0.3683 &	--86.0  &  153.5  \\
16 & 0.4909 &	--3.8  &  ---   \\
17 & 0.5172 &	17.7  &  ---   \\
18 & 0.5553 &	32.8  &  ----    \\
19 & 0.5915 &    62.7   &   --98.0  \\ 
19 & 0.6242 &	81.1  &   --120.5     \\
20 & 0.6374 &	82.7  &   --113.4    \\
21 & 0.6530 &	86.3  &   --136.7    \\
22 & 0.6551 &	88.9  &   --131.3   \\
23 & 0.6908 &   96.3 &   --156.1    \\
24 & 0.7115 &	99.8  &   --158.8     \\
25 & 0.7147 &	99.5  &   --159.4 \\
26 & 0.7319 &	105.7  &  --164.0\\
27 & 0.7407 &	107.4  &  --168.0\\
28 & 0.7670 &	109.6  &  --167.4\\
29 & 0.7754 &	109.2  &  --167.7\\
30 & 0.7950 &	102.6  &  --172.9\\
31 & 0.8112 &	111.7 &   --170.9 \\
32 & 0.8384 &	101.6  &  --169.0\\
33 & 0.8687 &	100.7 &  --150.6 \\
34 & 0.8978 &	94.2 &  --136.9\\
35 & 0.8989 &	92.0 &  --140.4\\
\hline \\
\end{tabular}
\end{center}
\end{table}

 The possibility of using cross-correlation techniques to
determine rvs for this type of spectrum was considered in 
the present authors' paper on the young early type close binary $\eta$ Mus (Budding et al., 2013).  The difference in results
for particular rv estimates, which could amount to several km s$^{-1}$, was explained in terms of sampling window arrangement, line asymmetries and high frequency noise.  The advantages
of fitting line-modelling functions for close binaries
with few and broadened features were discussed in some detail 
by Rucinski (2002).

We have used the same fitting model for the rv
variation, including regular proximity effects, as in Paper
I. Results of the application of this program to the observed
rvs are shown in Fig~7.
 
\begin{figure}
\label{fig-7}
\includegraphics[height=6cm,angle=0]{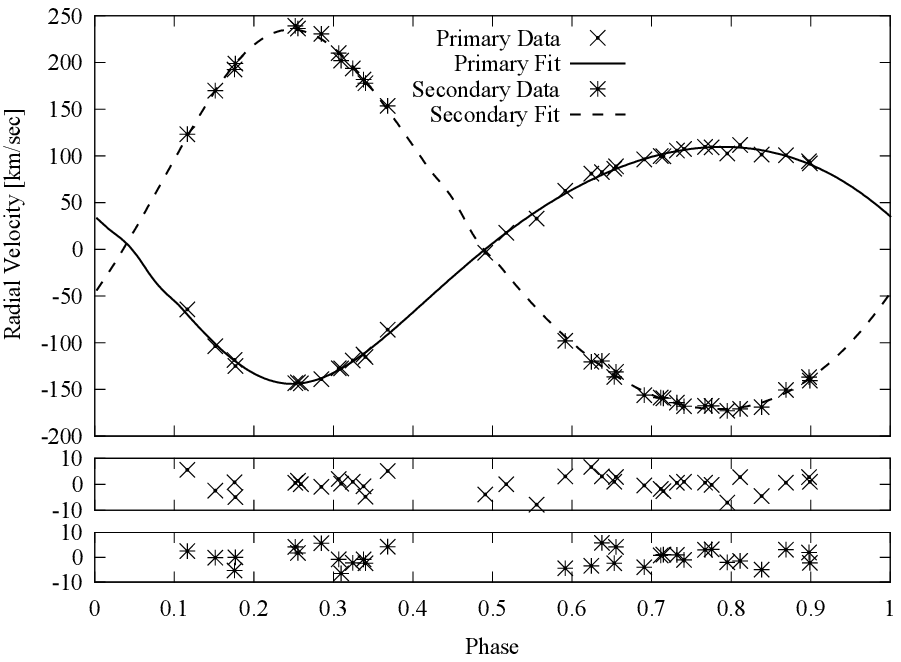}\\
\includegraphics[height=6cm,angle=0]{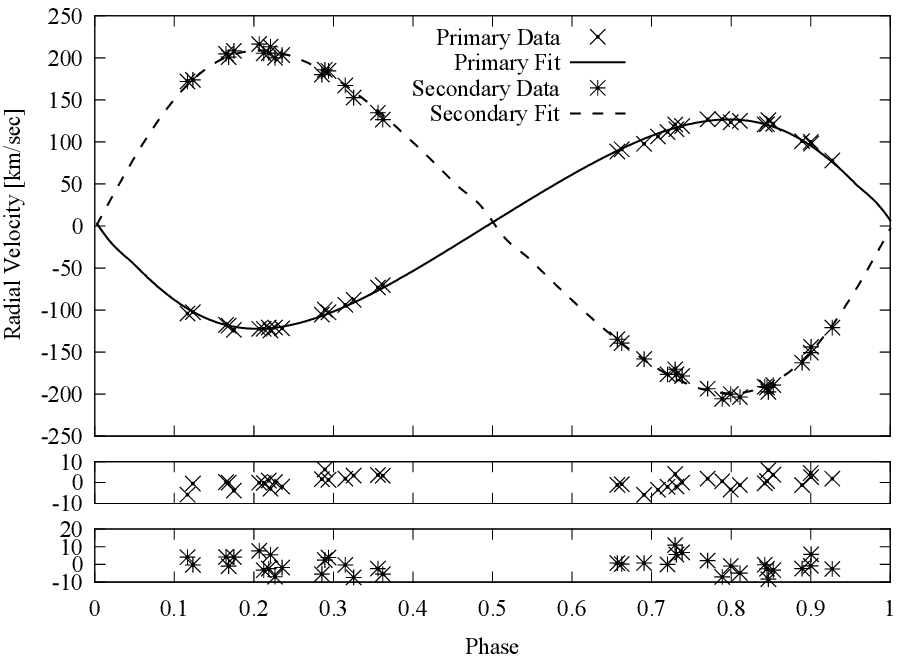}\\
\caption{The upper panel shows our best-fitting binary radial velocity curves for GG Lup,
with residuals plotted below. The residuals of the stronger primary lines
are less than those of the secondary.  The former appear consistent with the random scatter in individual line measures
of less than $\sim$6 km s$^{-1}$ amplitude. In the lower panel we have plotted a similar fitting to the data
of Andersen et al.\ (1993). The change in shape of the curves, due to the apsidal advance, can be noticed. The scatter in residuals appears comparable, in general, although
that for the primary may be somewhat less in the results of Andersen et al.\ (1993).}
\end{figure} 
 
 \subsection{Line profiles}
In view of the suspected oscillatory behaviour discussed above,
we sought to check for evidence of this
in the line profiles (see e.g.\ De Cat, 2002, for a review).
Using the numerical model's radial velocity parameters 
given in the previous subsection, the spectral response 
covering an 8 \AA\ range around the HeI 6678 feature was shifted to the primary star's reference frame.
Results are shown in Fig~8.
An additional distorting effect on the primary profile  corresponding
to an eclipse phase may be seen at the top of this figure.
However, any irregularity in the primary line's behaviour through
the rest of the phase range may be regarded as intrinsic.

Frequency decomposition was calculated for the wavelength window
shown in Fig~8.  The amplitudes and phases of a few of the best-fitting
lower frequency contributions were considered
in relation to their possible dependence on orbital phase.
An anomaly near orbital phase 0.1 was noticed,
and this might relate to the `heartbeat' effect noted in Fig 6.
But otherwise, the amplitudes showed no clearly discernible 
pattern, against their sizeable inherent scatter, 
that could be associated with a multiple
of the orbital frequency.  The phases of these
low frequency contributions similarly suggested
that the line generally tends to remain predominantly 
symmetrical about its centre.
We thus found little clear evidence of any systematic
frequency-dependent behaviour in these line profiles.
Andersen et al.\ (1993) indicated that some of 
the lines they examined were not consistent 
with the expected trend, without expanding on
possible causes.

As a check on our decomposition, 
the reverse transformation has been given in Fig~8
as dashed lines for each measured orbital phase. 
Any regular variations of oscillation 
with the orbital phase should become visible in that way.
But the data-quality, in the present set of spectrograms,
appears of insufficiently high S/N quality to enable a detailed analysis
of line profile variations with reliability.  Nevertheless, 
superficial indications of non-constancy to the profiles of the primary (in particular), merit follow-up research. 
 
\subsection{Rotational velocities}
\begin{figure}
\label{fig-8}
\includegraphics[height=11cm,angle=0]{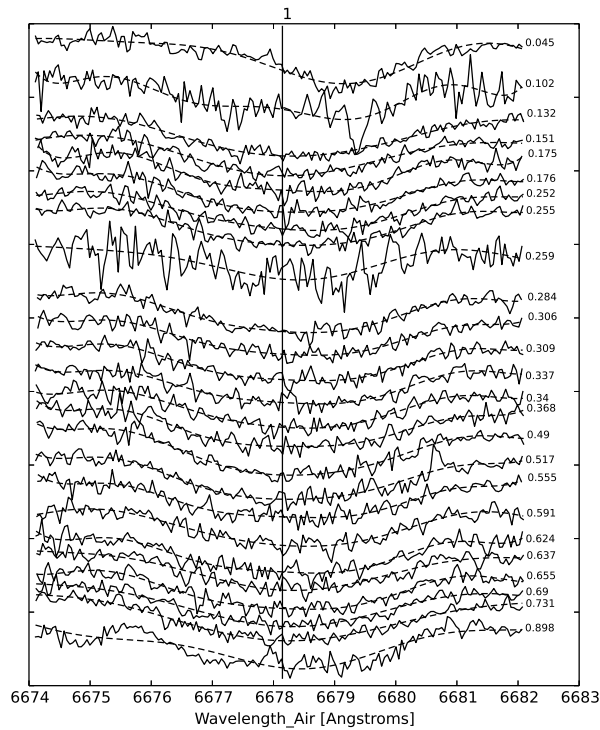}\\
\caption{The series of He I profiles are labelled on the right with
their orbital phase. The vertical spacing has been adjusted to allow feature visibility.
The vertical tick marks are spaced at intervals of 10\% of the mean continuum.
In this way it can be seen that central line depths reach to typically
$\sim$5\% of the continuum. 
}
\end{figure}


We consider rotation speed measures in the context of the 
eccentric orbit, for which, following classic expositions of 
spectroscopic binaries (e.g.\ Smart, 1960), the primary's mean orbital velocity satisfies
\begin{equation}
\overline{v}_{1{\rm orb}}= na_1 = \frac{(\alpha + \beta)_1(1 - e^2)^{1/2}}{2 
\sin i} \,\,\,   ,
\end{equation}
where $\alpha$ and $\beta$ are the rv excursions above and below the mean, and other symbols have their usual meanings
(with a corresponding form for the secondary).
The primary and secondary have separations from the centre of mass in the ratio of 0.388, 0.612, respectively, so the mean radii will take up 0.518 and 0.243 of these separations.
This implies mean values for the  equatorial rotation speeds of 65.2 km s$^{-1}$ and 48.2 km s$^{-1}$ 
for primary and secondary respectively.

 For comparison purposes, a useful direct estimate of the synchronous rotation velocity
for uniformly rotating spherical stars in circular orbits can be shown to be given by $v_{\rm sync} = 50.6 R/P$, 
where $R$ is the stellar radius in solar units and $P$ is the orbital period in days.  This would yield 
66.2 and 49.0 km s$^{-1}$ for primary and secondary values of $v_{\rm sync}$.

As a consequence of elliptic motion, the total orbital velocity
(one component about the other) undergoes a variation of the form:
\begin{equation}
v_{\rm orb} = \overline{v}_{\rm orb} \sqrt{\left( \frac{1 + 2e\cos\nu + e^2}{1 - e^2} \right)} \,\,\,   ,
\end{equation}
 hence, at periastron, the velocity is enhanced by the factor $\sqrt{(1+e)/(1-e)}$ = 1.175.  Not only that, but since the separations
 from the centre of gravity are then reduced by the factor $(1-e)$ on the mean, synchronous rotation speeds at periastron
would increase to about 91.2 and 67.4 km s$^{-1}$, for primary and secondary respectively.

\begin{figure}
\label{fig_rotation}
\label{fig-9}
\begin{center}$
 \begin{tabular}{@{}l@{}l}    
 \hspace{-0.25cm}
 \includegraphics[width=4.2 cm,height=3.4 cm]{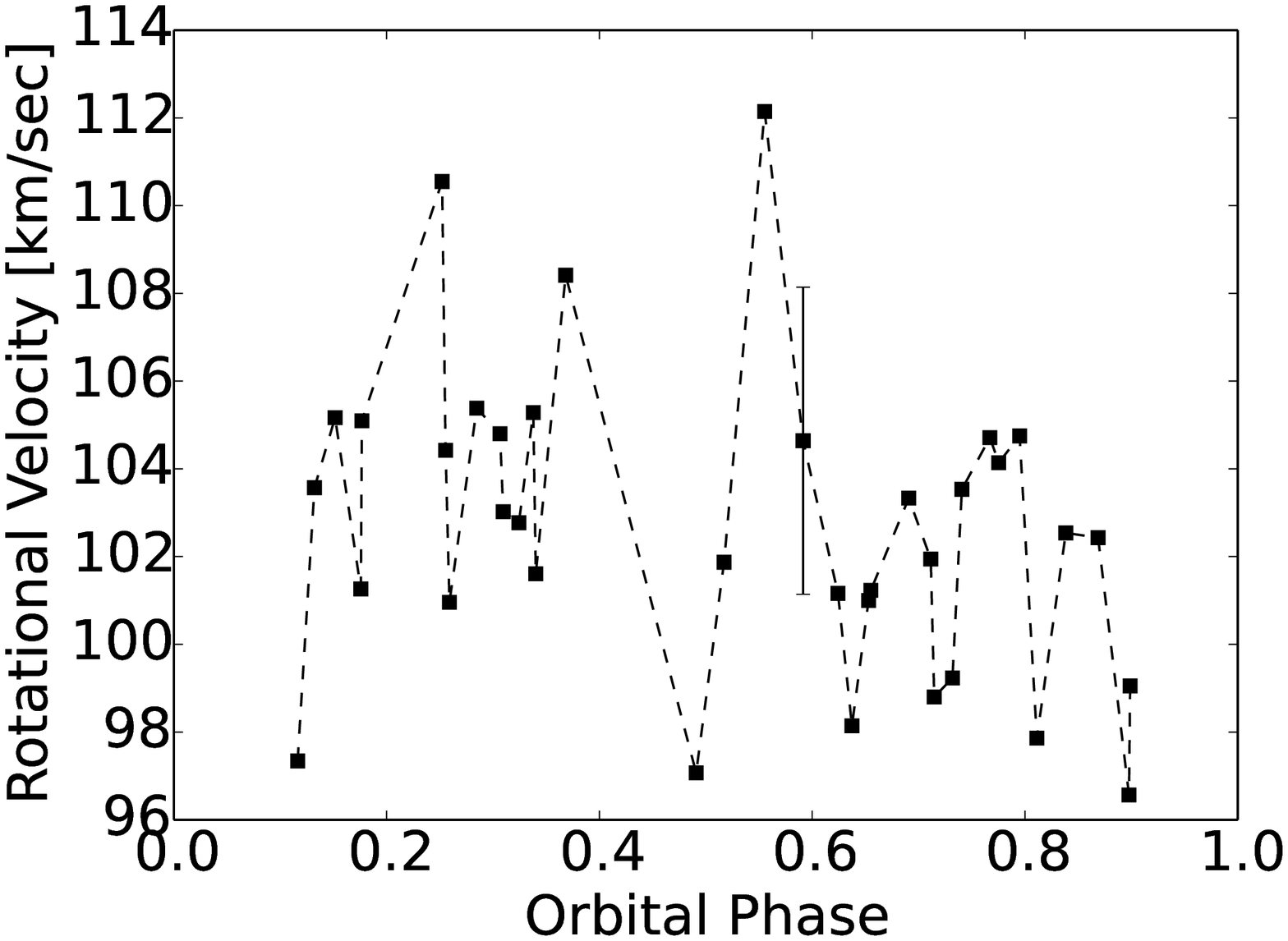} &
 \includegraphics[width=4.2 cm,height=3.4 cm]{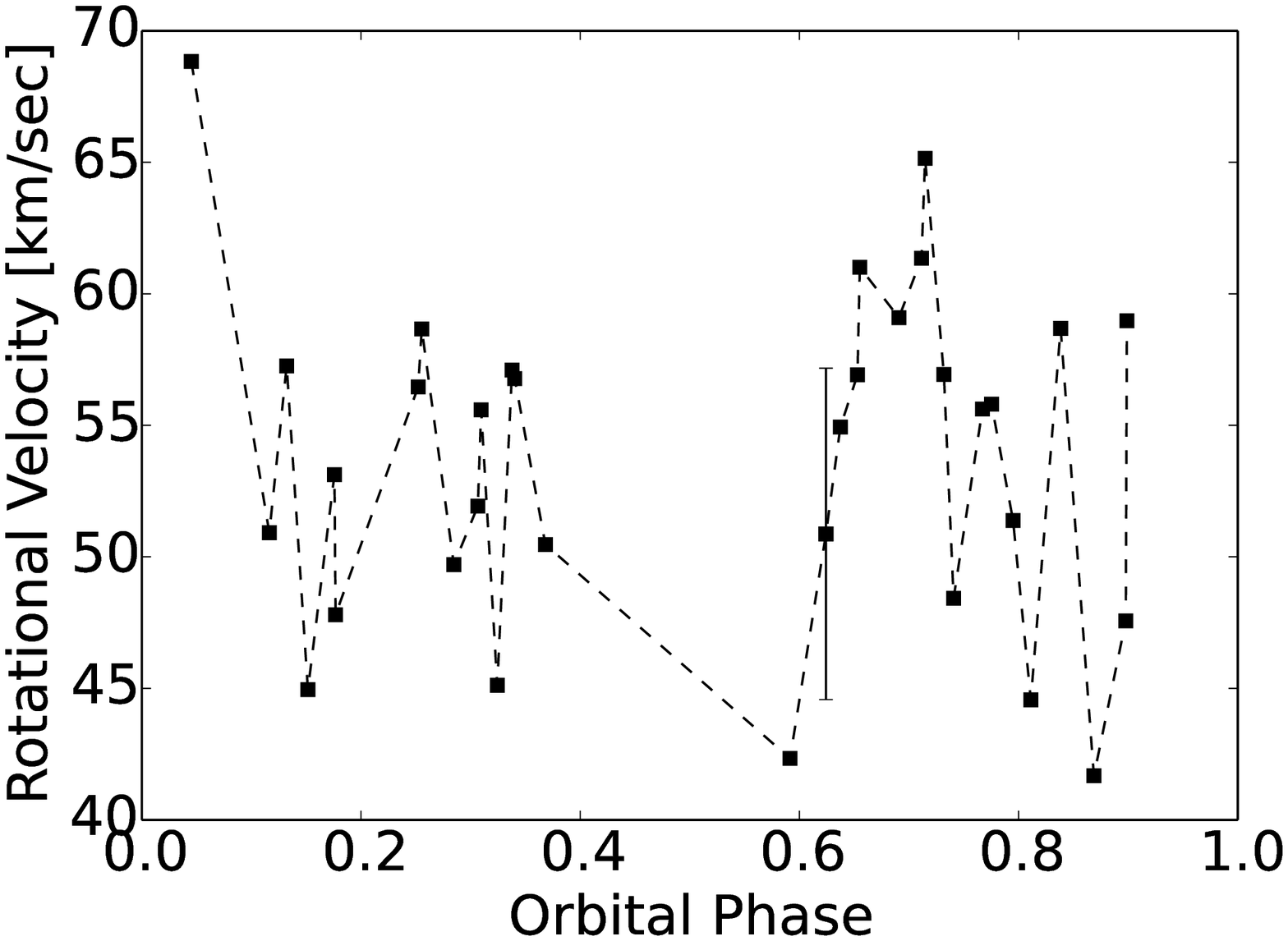} \\
 \end{tabular}$
 \caption{Primary and secondary rotational velocities for GG Lup.
 Due to the Rossiter-McLaughlin Effect the interpretation of values close to phases 0.0 and 0.5 will be uncertain.
}
 \end{center}
 \end{figure}

 Measured rotation speeds are shown in Fig~9.
 The mean apparent rotation speed of the primary at 102.6$\pm$9.5 km 
 s$^{-1}$ is not far from a 3/2 resonance with the value corresponding to synchronism with the mean orbital speed, but it is still some 15\% greater
 than locking at the periastron.  This 
 has to be seen against the complications already 
 mentioned, associated with (non-radial) pulsational effects.
 These imply not only that the measured line widths 
 may arise from effects additional to the simple rotation
 of a near-spherical body, but that the additional role
 of a $\kappa$-mechanism may sustain motions that would
 otherwise become damped out.  The scatter of results in 
 Fig~9 impede a clear view of any 
 such possible phase-dependent effects, though the 
 rotation of the primary appears generally greater in the
 first half of the orbital cycle than the second.
 
 For the secondary, however, the mean rotation speed of 55.3$\pm$7.2 km s$^{-1}$ appears in a more expectable range, i.e.\ enhanced from the mean synchronization, due to greater
tidal interaction near periastron, but not at the highest
possible value in that respect.

  
 
\subsection{Absolute parameters}
Absolute parameters for GG Lup, derived from combining the 
foregoing photometric and spectroscopic results and Kepler's third law,
are listed in Table~7. The mean photometric parallax (Eqn 3.42 in Budding \& Demircan, 2007) of 155 pc is within a reasonable error estimate of 20 pc
to the astrometric value 167 pc given by HIPPARCOS. 

 The absolute parameters of Table~7 are close to those of Andersen et al.\ (1993)\footnote{{\bf Andersen et al.\ (1993) gave component masses of 4.12$\pm$0.04 and 2.51$\pm$0.02 M$_{\odot}$, and radii of 2.38$\pm$0.03 and 1.72$\pm$0.02 R$_{\odot}$.}},  particularly for the primary component, where the mass and radius values come well within our own error
limits, and almost within those given by Andersen et al.  Even for the secondary, whose properties are less
well-determined, the differences in mass and radius are almost within our error measures.
Given that the masses depend on the cube of size
determinations, however, the very low error estimates of Andersen et al.\ appear questionable. 
Andersen et al.\ (1993) laid weight on these parameters in making inferences about the
composition of young early type stars like those of GG Lupi, which, in turn, have
interesting implications for the origin of Gould's Belt stars.  This point is
taken up in Section 4, below.

\begin{table}
\begin{center}
\caption{Adopted absolute parameters for the GG Lup system. 
Formal errors are shown by parenthesized numbers to the right 
and relate to the least significant digits in the corresponding parameter
values.}
\label{tbl-7}
{\footnotesize
\begin{tabular}{lr}
\hline
\multicolumn{1}{c}{Parameter} & \multicolumn{1}{c}{Value} \\
\hline
Period (days) & 1.8495927  \\
Epoch (HJD) & 2448349.0625  \\
Mags. $V$, $(B - V)$, $(V - J)$  & 5.589; --0.099; --0.530\\
Vel.\ amplitudes $K_{1,2}$ (km/s)& 127.5(1.0); 201.0(1.6) \\
Star separation $A$ (R$_{\odot}$)& 12.01(0.09) \\
System vel. $V_{\gamma}$ (km/s)& 4.0(1.0) \\ 
Masses $M_{1,2}$ (M$_{\odot}$)& 4.16(0.12); 2.64(0.12)\\
Radii $R_{1,2}$  (R$_{\odot}$) & 2.42(0.05); 1.79(0.04) \\
Surface grav. $g$ (log cgs) & 4.28(0.04); 4.30(0.05) \\
Prim.\ mag. $V_{1}$ & 5.84(0.03) \\ 
Sec.\ mag. $V_{2}$ & 7.29(0.03) \\ 
Prim.\ temp. $T_{e,1}$ (K)& 13000(300) \\ 
Sec.\ temp. $T_{e,2}$ (K)& 10600(300) \\ 
Distance &  160(20) pc \\
\hline
\end{tabular}
}
\end{center}
\end{table}

\section{$\mu^1$ Sco}

The bright and hot young close binary $\mu^1$ Sco (= HD 151890, HIP 82514; $V \approx$ 2.98, $B - V \approx$ --0.16, $U - B \approx$ --0.85, $V - J \approx$ --0.48, $V - H \approx$ --0.50, $V - K \approx$ --0.72) was found to be a spectroscopic binary
already over a century ago by S.\ I.\ Bailey (cf.\ Pickering, 1896).  
Its membership of the Sco-Cen OB2 Association (Blaauw, 1964; Preibisch \& Mamajek, 2008) has been adopted by many previous authors, who have used a corresponding age 
estimate to constrain modelling.
The sky location RA(2000) = 16h 51m 52.23s; dec(2000) = --38\degr 02\arcmin 50.6 \arcsec; distance $\sim$160 $\pm$ 20 pc and proper
motions ($\mu_{\alpha} \cos \delta \approx -11$, $\mu_{\delta} \approx -22$ mas y$^{-1}$) imply that the star is now located several degrees east of the main
 UCL concentration of the association on the sky, just $\sim$4\degr\ from the 
 central plane of the Disk.
 
As with GG Lup, it was quite a long time after the original binary identification when Rudnick and Elvey (1938) demonstrated
eclipses, using an early type of photoelectric photometer.  
This was quickly followed by 
van Gent's (1939) photographic light curve.  
  The history of observations of the binary was recently reviewed by
van Antwerpen and Moon (2010; hereafter vAM), whose Figure 1 gives an interesting montage of 7 light curves published
between 1938 and 2009.  The typical scatter in this montage is a few hundredths of a 
magnitude. These might be understood in terms of the
available accuracies of older generation light detectors,
particularly photographic ones.  The phase range after the secondary minimum to around the second quadrature shows
a larger than normal scatter, however, suggestive of some inherent additional light fluctuations.

The scatter about the fitting to the HIPPARCOS light curve (ESA, 1997)
is generally less than that which appears in vAM's Fig~1,
but it is still appreciably larger than typical  ($\sim$0.01 mag, 
see our Fig~1, for example), and it may have some systematic trends, more noticeably
 in the first half of the phase cycle in this data-set.

\begin{figure}
\label{fig-10}
\noindent\includegraphics[height=8cm,angle=-90]{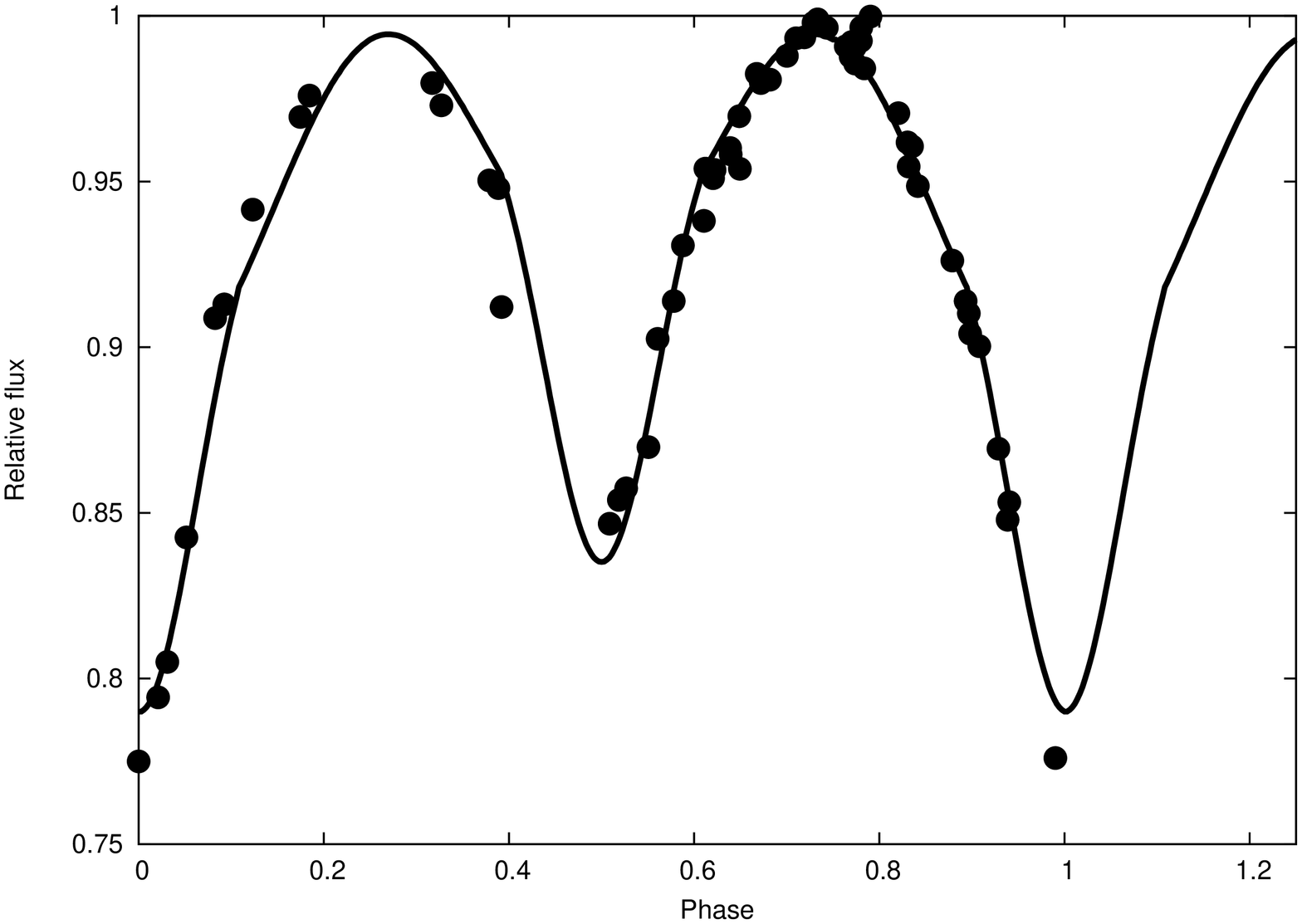}\\
\noindent\includegraphics[height=8cm,angle=-90]{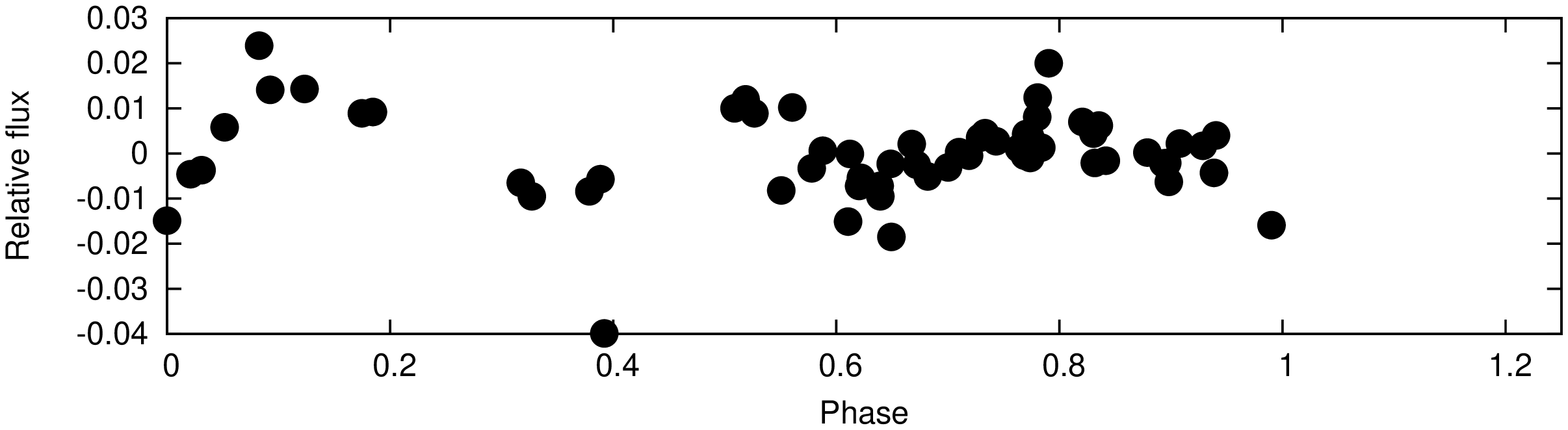}\\
\caption{HIPPARCOS V photometry of $\mu^1$ Sco and model fitting
with residuals below (see Section 4.1 for details).}
\end{figure}

\subsection{Photometry and analysis}
The uniformly obtained HIPPARCOS
data (ESA, 1997), was published 
with the ephemeris
 Min I = 2448500.1010 + 1.44626E.
 This may be compared with that of vAM
 who gave 
 Min I = 2449534.1770 + 1.4462700E, after combining several 
light curves spanning a 70 y interval. 
There appeared no definite evidence for any
systematic variation of period beyond the sixth 
digit after the decimal point (in units of d)
clearly distinguished in previous literature,
so we adopted the light elements of
vAM for our analysis.  In Figure~11
we show recent relatively high accuracy times of minima obtained
within the southern binaries programme of the RASNZ VSS
that support this point, although with some
suggestion of a slight tendency to
period increase. 

\begin{figure}
\label{fig-11}
\hspace{1em}
\noindent\includegraphics[height=5cm]{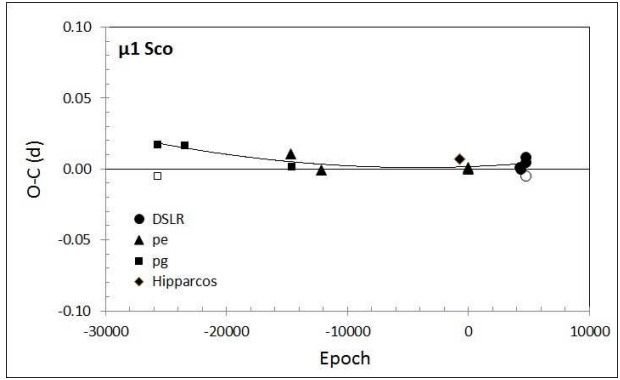}\\
\caption{Times of minima for $\mu^1$ Sco (primary, full symbols;
secondary, open).
We have added our own recent DSLR timings
to the historic ones reported by vAM.
There is a suggestion of a slight period increase
indicated by the parabolic fit, but the second-order term in
the epoch number is less than 10$^{-10}$ and clearly low
when placed alongside comparable Algol systems 
(Kreiner and Zi\'{o}{\l}kowski, 1978). }
\end{figure}

The optimized model shown in Fig~12, together with the
averaged and normalized light curve of vAM,
corresponds to the parameters listed to the right in Table~8.
The temperatures in this table are influenced by the
values considered by vAM, as also the adopted mass-ratio.
vAM argued that the photometric data would allow appreciable variation in the assigned temperatures.
On the other hand, Sahade and Garcia's (1987)
fitting of the UV continua appeared to give
a relatively firm determination of the primary
temperature (25000 K). 

\begin{table}
\begin{center}
\caption{Curve fitting results for $V$ photometry   
of $\mu^1$ Sco.  Error estimates are given as
bracketed numbers referring to the last significant digit.
\label{tbl-8}} 
\begin{tabular}{lcc}
   \hline  \\
\multicolumn{1}{c}{Parameter}  & \multicolumn{2}{c}{Value} \\
\multicolumn{1}{c}{ } & HIP & vAM \\
\hline \\
$T_h$ (K) & 24000  & 24000 \\
$T_c$ (K) & 17000  & 17000 \\
$M_2/M_1$ & 0.627  & 0.627  \\
$L_1$ & 0.48(7) & 0.55(6) \\
$L_2$ & 0.52(7) & 0.45(6) \\
$r_1$  & 0.350(6)& 0.311(2) \\
$r_2$  & 0.373(8)& 0.366(3) \\
$i$ (deg) & 62.6(8) & 64.4(3)\\
$e$  & 0.0 & 0.0\\
$u_1$ & 0.24 & 0.24\\
$u_2$ & 0.30 &0.30\\
$\Delta\phi_0$(deg) & 0.0 & 0.0 \\
$\chi^2/\nu$ & 1.03 & 0.94 \\
$\Delta l $& 0.01 & 0.007 \\
\hline
\end{tabular}
\end{center}
\end{table}

Although the mean (s.d.) dispersion of 0.01 mag appears about typical
for a HIPPARCOS data-set, here with only 62 points, the scatter seems 
not so uniformly distributed, echoing suggestions from earlier observers of an intrinsic variability beyond that associated with a regular eclipsing binary system.  The two fittings of Table~8 are appreciably different,
in comparison with the formal error estimates, suggesting some
lack of modelling adequacy to cover the inherent complexity of the system.

In agreement with previous authors is our initial result that the secondary star is larger than the primary, and certainly larger than
a normal Main Sequence relationship would imply for the given primary 
radius and  the secondary's much lower mass, that we confirm later.
The system must therefore be in a state involving binary  
(i.e.\ interactive) evolution.
  The relatively large 
masses, high primary surface temperature, and proximity of the components make this a very atypical arrangement 
for a `classical' Algol binary, however.  The similarity of the period and mass-ratio to that of GG Lup may be noticed, though it quickly becomes clear 
that $\mu^1$ Sco must be in quite a different condition.
The marked proximity effects of the light curve in Fig~12 
cannot be explained by two near Main Sequence stars.

\begin{figure}
\label{fig-12}
\noindent\includegraphics[height=8cm,angle=-90]{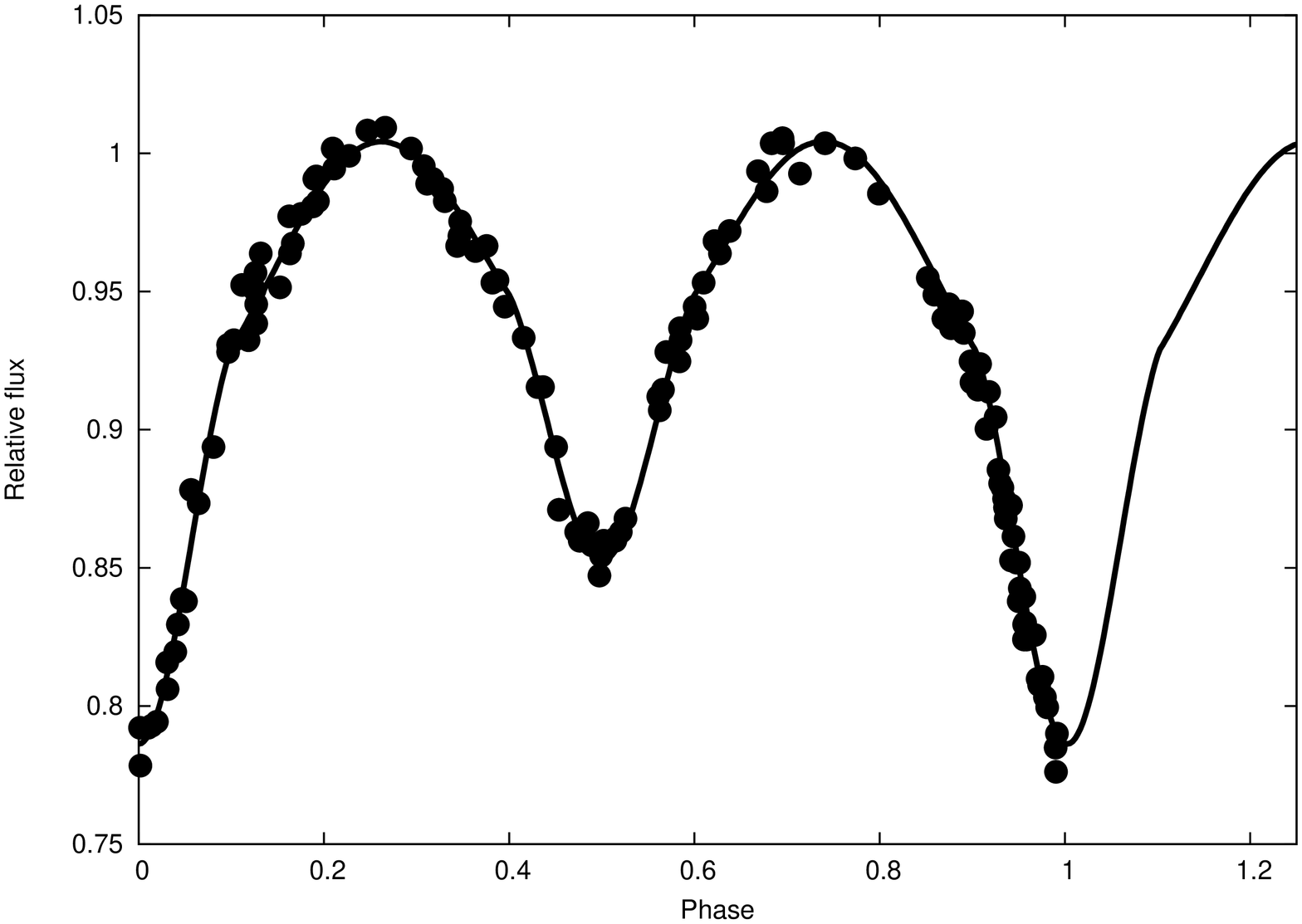}
\noindent\includegraphics[height=8cm,angle=-90]{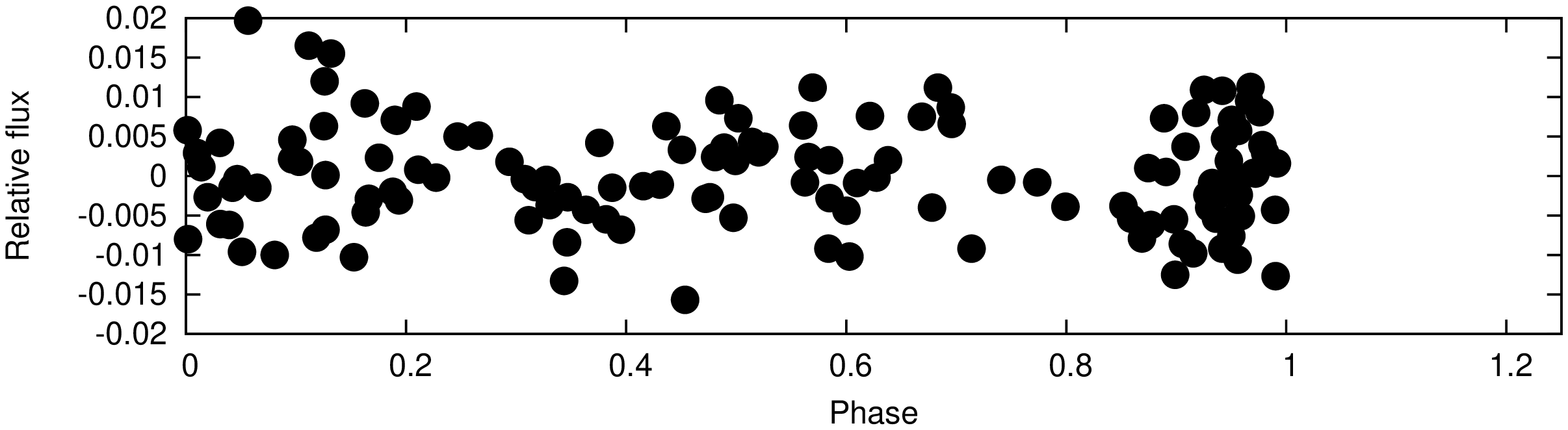}\\

\caption{Normalized V-band photometry of $\mu^1$ Sco, after van Antwerpen and Moon (2010),
and model fitting.}
\end{figure}

The scatter of individual light curves, like that shown in Fig~10, 
and the historic ones reviewed by vAM can be considerably reduced by averaging (`normalizing') data over short phase ranges.  In this way, the data-set shown in Fig~12 was produced
by vAM, and analysed to find parameters that
are similar to those we have found and listed in Table~8. Our
secondary relative radius appears  larger than that of vAM,
but otherwise the parameters (ours and vAM's) are
generally comparable.

 $BVR$ photometry of selected eclipsing binaries of the southern sky
 by members of the VSS section of the RASNZ (cf.\ Section 2.1)
  has included 
 $\mu^1$ Sco, using the same (DSLR) equipment.
 The light curves are shown in Fig~13, where we also show optimal fits to the reduced $B$, $V$ and $R$ datasets and the $V$ residuals.
The parameters going with these fits are listed in Table~9, where similarity of the fitting to the collected literature data of
vAM (Table~8, right) may be noticed.  
The fractional luminosities  of the two components ($L$)
in the optical region can be seen to be comparable.  This also follows from the temperature and radius ratios,
given that the $B V R$ filters fall into the `Rayleigh-Jeans' radiation regime (approximately). The slight preference in the results for $L_1 > L_2$ is suggestive
that the assigned value of $T_c$ may be slightly too high.

\begin{table}
\begin{center}
\caption{Curve fitting results for this paper's $BVR$ photometry 
of $\mu^1$ Sco.
\label{tbl-9}} 
\begin{tabular}{lccc}
   \hline  \\
\multicolumn{1}{c}{Parameter}  & \multicolumn{3}{c}{Value} \\
\multicolumn{1}{c}{ } & $B$ & $V$ & $R$ \\
\hline \\
$T_h$ (K) & 24000  &  & \\
$T_c$ (K) & 17000  &  & \\
$M_2/M_1$ & 0.55  &  &  \\
$L_1$ & 0.52(3) & 0.52(3) & 0.52(4)   \\
$L_2$ & 0.48(3) & 0.48(3) & 0.48(4)  \\
$r_1$  & 0.311(2) & &  \\
$r_2$  & 0.369(3) & &   \\
$i$ (deg) & 64.3(3)  &  & \\
$u_1$ & 0.29 & 0.25 &0.21\\
$u_2$ & 0.35 &0.30 &0.25 \\
$\Delta\phi_0$(deg) & 0.0 & 0.0 & 0.0\\
$\chi^2/\nu$ & 1.03 & 1.01 & 1.02\\
$\Delta l $& 0.006 & 0.0055 & 0.007 \\
\hline
\end{tabular}
\end{center}
\end{table}

\begin{figure}
\label{fig-13}
\noindent\includegraphics[height=9cm,angle=-90]{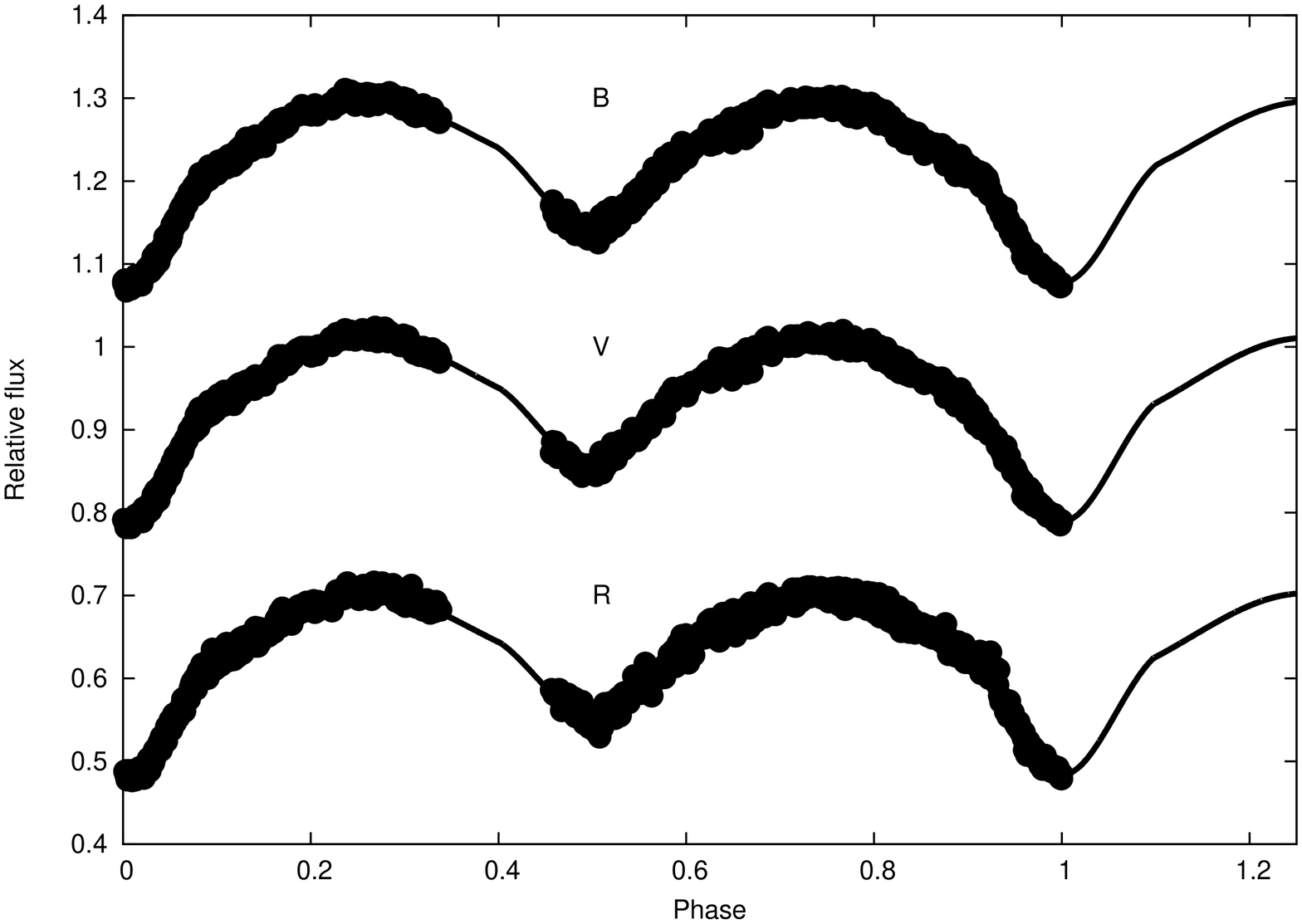}\\
\noindent\includegraphics[height=9cm,angle=-90]{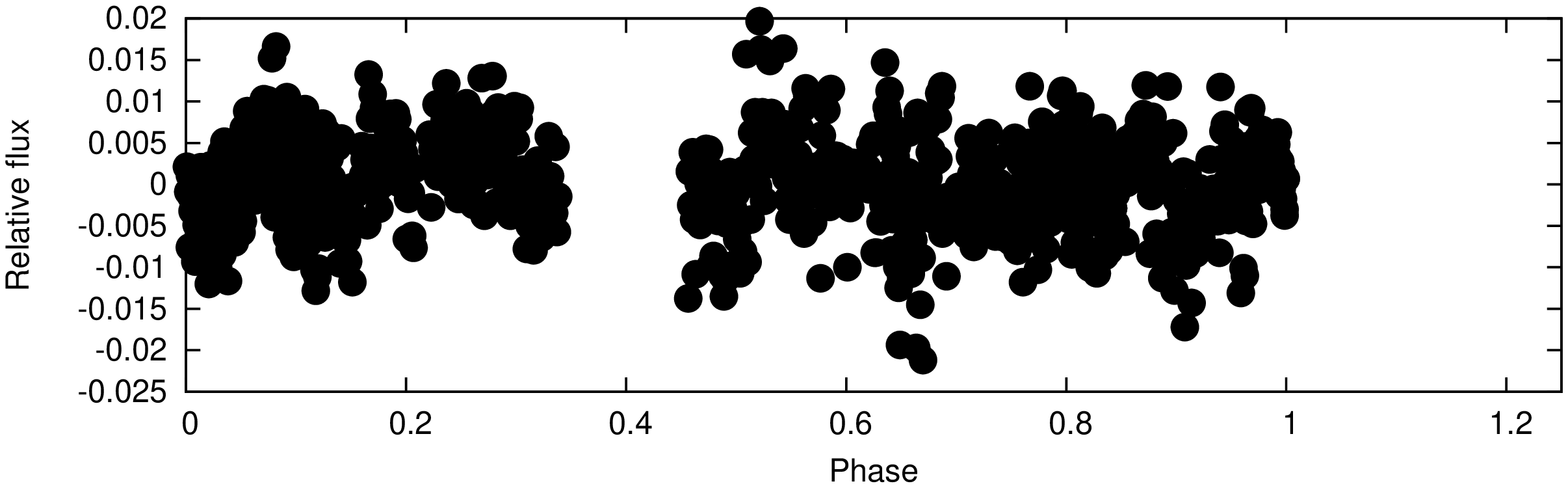}\\
\caption{Radau model fitting (Rhodes \& Budding, 2014) to DSLR $BVR$ light curves of $\mu^1$ Sco.
The residuals from the $V$ fitting are shown below.}
\end{figure}

\subsection{Spectroscopy} 
Our spectroscopic data were gathered using the
HERCULES spectrograph with arrangements essentially
similar to those described in Section 2.2.
{\bf The data were gathered over 3 allocations at MJUO in May 2006,
August 2010 and August 2011.
A new  SI600s type camera replaced
the SITe one used for the 2006 data.
The high brightness of $\mu^1$ Sco entailed
that exposures could be taken in relatively poor weather.
Seeing conditions would therefore often be relatively
poor and the S/N ratio
for several of the 24 spectrograms is not 
as high as the brightness of the star might suggest.
In fact,  when recording spectral images with  HERCULES it is normal
to allow the integrated photon count, which is displayed as the
exposure time ($t$) proceeds, to be terminated at a value corresponding to
a continuum S/N of $\sim$100 for the lower used orders: typically $\sim$50-200 s (cf.\ Section 2.2).

A compromise is involved here in that while the S/N varies in proportion 
only to $\sqrt{t}$, longer exposures for binaries of shorter period
will smear out the Doppler shifting and lose velocity resolution.
The rotational broadenings for  $\mu^1$ Sco are about double (primary)
or treble (secondary) those of GG Lup however (Section 3.4).  The precision
with which line positions can be specified is correspondingly reduced,
as is reflected in the dispersion panels of Figure 14.}
The data were reduced using the (revised) software {\sc hrsp} 5
(Skuljan, 2012), further analysis being then carried out using {\sc iraf}.

\subsection{Radial velocities}

\begin{table}
\begin{center}
\caption{Spectral lines of $\mu^1$ Sco which were used for radial velocity determination.
\label{tbl-10}}
\begin{tabular}{ccll}
  & & &  \\
\multicolumn{1}{c}{Species}  & \multicolumn{1}{c}{Order no.} &
\multicolumn{1}{l}{Wavelength} & \multicolumn{1}{l}{Comment}  \\
He I & 85        &  6678.149  &   strong p,  moderate s       \\
He I &  97       &  5875.65  &  strong p, usable s\\
He I & 113 		&   5015.67 & weak p, very weak s \\
He I & 115 		& 4921.93 & moderate p, weak s \\
He I & 121       &  4713.201  &  weak p, weak s  \\
\hline \\
 
\end{tabular}
\end{center}
\end{table}

 The central issue of locating lines for rapidly rotating
early type stars was discussed in Sections 2.2 and 2.3 above.
Additional complications associated with the ongoing interactive
evolution of the system are likely to be also present in the case
of  $\mu^1$ Sco.
Profile fitting using  `dish shaped' functions convolved with Gaussians 
set against slightly sloping continua was attempted, but the 
large broadening of the generally not intrinsically
strong and possibly distorted features resulted in relatively poor determination of positions. The
 lines listed in Table~10 provide a guide
 to what was dealt with, but in many cases only the He I lines
 allow shifts to be determined with an acceptable accuracy of
 several km s$^{-1}$.

The final set of rv measurements, which were derived from the fitting function's centroid positions, is given in Table~11.
  Dates and velocities and have been corrected
to heliocentric values as before and are shown in Fig~14. 
 Although the rvs are again listed with one significant
decimal place in Table~11, this gives an over-optimistic
impression of the accuracy of the determination (cf.\ Section 2.3). 

\begin{table}
\begin{center}
\caption{Radial velocity data for $\mu^1$ Sco.
\label{tbl-11}} 
   \begin{tabular}{cccc} 
\multicolumn{1}{c}{No}  & \multicolumn{1}{c}{Phase} &
\multicolumn{1}{c}{RV1} & \multicolumn{1}{c}{RV2} \\
\multicolumn{1}{c}{}  & \multicolumn{1}{c}{} &
\multicolumn{1}{c}{km s${-1}$} & \multicolumn{1}{c}{km s$^{-1}$} \\
\hline \\
 1 & 0.0041 &      8.5   &   1.7  \\
 2  & 0.0070 &  ---       &  16.2 \\
 3  & 0.0097 &  ---       &  14.4 \\ 
 4 & 0.0445 &	--56.9   &  90.3 \\ 
 5 & 0.0496 &	--81.8   &  72.3 \\
 6 & 0.0501 &	--64.1   &  92.0  \\
 7 & 0.0525 &	--79.7   &  76.6 \\
 8 & 0.2313 &	--143.4  &  243.1 \\
 9 & 0.2615 &	--147.8  &  244.2  \\
10 & 0.3087 &	--150.4  &  203.1  \\
11 & 0.3116 &	--140.8  &  218.5   \\
12 & 0.3524 &	--113.4  &  214.4   \\
13 & 0.3682 &	--96.4   &  205.1   \\
14 & 0.3701 &	--101.6  &  203.8   \\
15 & 0.6114 &	  102.8  & --152.6   \\
16 & 0.6141 &	  104.6  & --150.6 \\
17 & 0.6143 &	   80.9  & --179.3  \\
18 & 0.6466 &	   95.2  &  --229.1   \\
19 & 0.6953 &	  118.3  &  --237.5   \\
20 & 0.6981 &	  123.4  &  --245.6    \\
21 & 0.7217 &     108.5  &   --278.3  \\ 
22 & 0.8470 &	  119.8  &   --217.0     \\
23 & 0.9887 &	  --1.1  &   ---    \\
24 & 0.9914 &	  --3.2  &   ---    \\
\hline \\
\end{tabular}
\end{center}
\end{table}

\begin{figure}
\label{fig-14}
\includegraphics[height=6.5cm,angle=0]{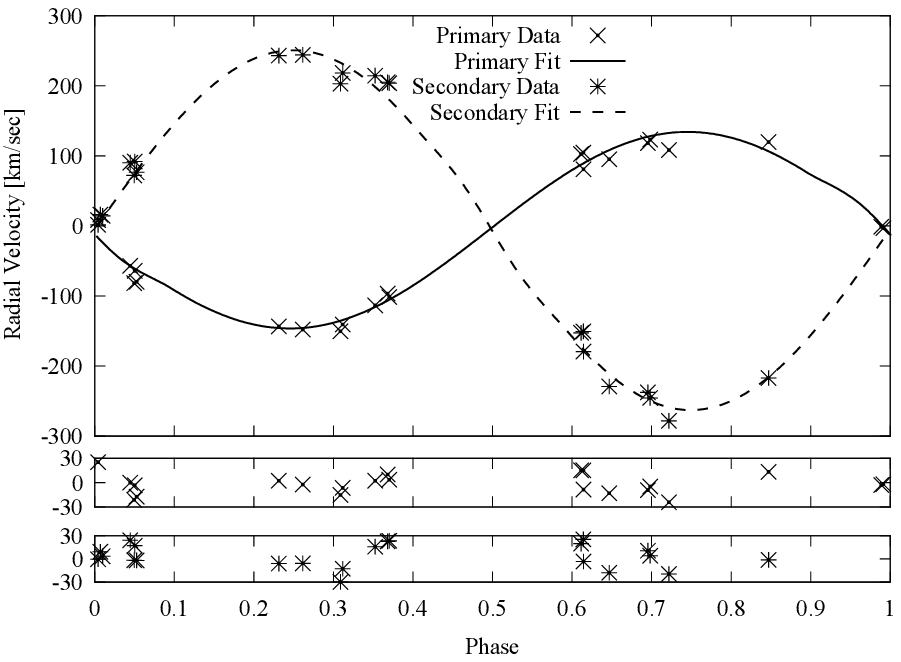}\\
\includegraphics[height=6.5cm,angle=0]{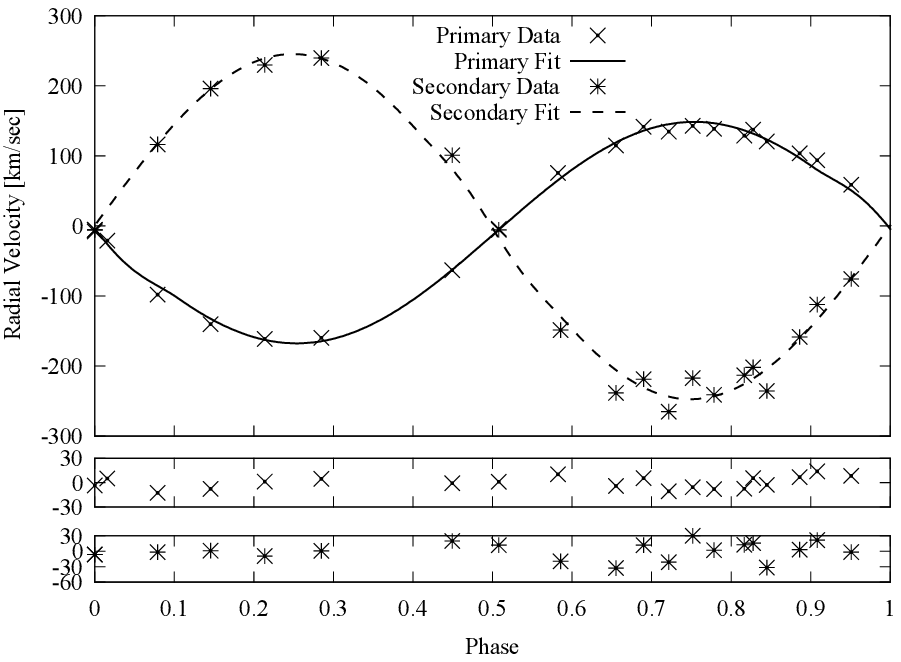}\\
 
\caption{The upper panel shows the best fitting binary rv curves for our
 observations of $\mu^1$ Sco, 
with the residuals plotted below. The excursions of the stronger primary lines are slightly less than those of the secondary.   The former appear consistent with the random scatter in the individual measures of less $\sim$12 km s$^{-1}$ amplitude. The lower panel shows
our fitting to the rv data assembled by vAM (2010). Again
we notice a comparable scatter in the residuals, relating to the
difficulties of locating the few, very broad and shallow lines.}

\end{figure}

\subsection{Rotation velocities}
Measured rotational speeds are shown in Fig~15. The mean apparent rotation speed of the primary at 191.5  km s$^{-1}$ is faster than synchronization by a factor of approximately 1.5.
For the secondary however, the mean rotation speed of 165.0  km s$^{-1}$ appears closer to synchronization, though still $\sim$10\% higher.

 The high rotation speed of the primary is again an issue, since the
synchronization timescale is expected to be fairly short in comparison 
to the Main Sequence time for close pairs, as noted in Section 2.1.
The circumstance of ongoing interactive evolution
will significantly complicate the angular momentum history, 
however, so that it is not obvious that standard models can be applied to 
$\mu^1$ Sco.

\begin{figure}
\label{fig-15}
\begin{center}$
 \begin{tabular}{@{}l@{}l}    
 \hspace{-0.25cm}
 \includegraphics[width=4.2 cm,height=3.4 cm]{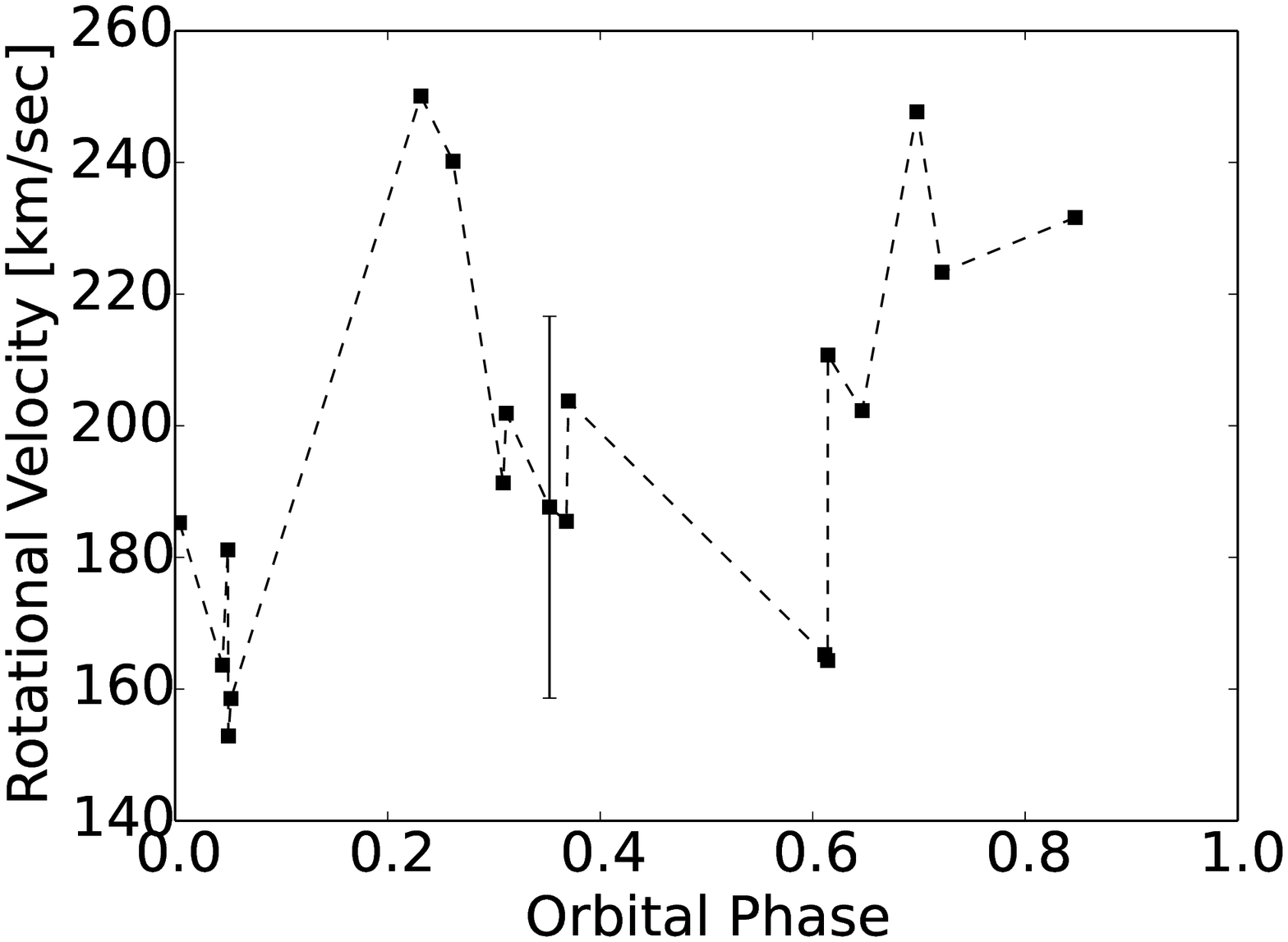} &
 \includegraphics[width=4.2 cm,height=3.4 cm]{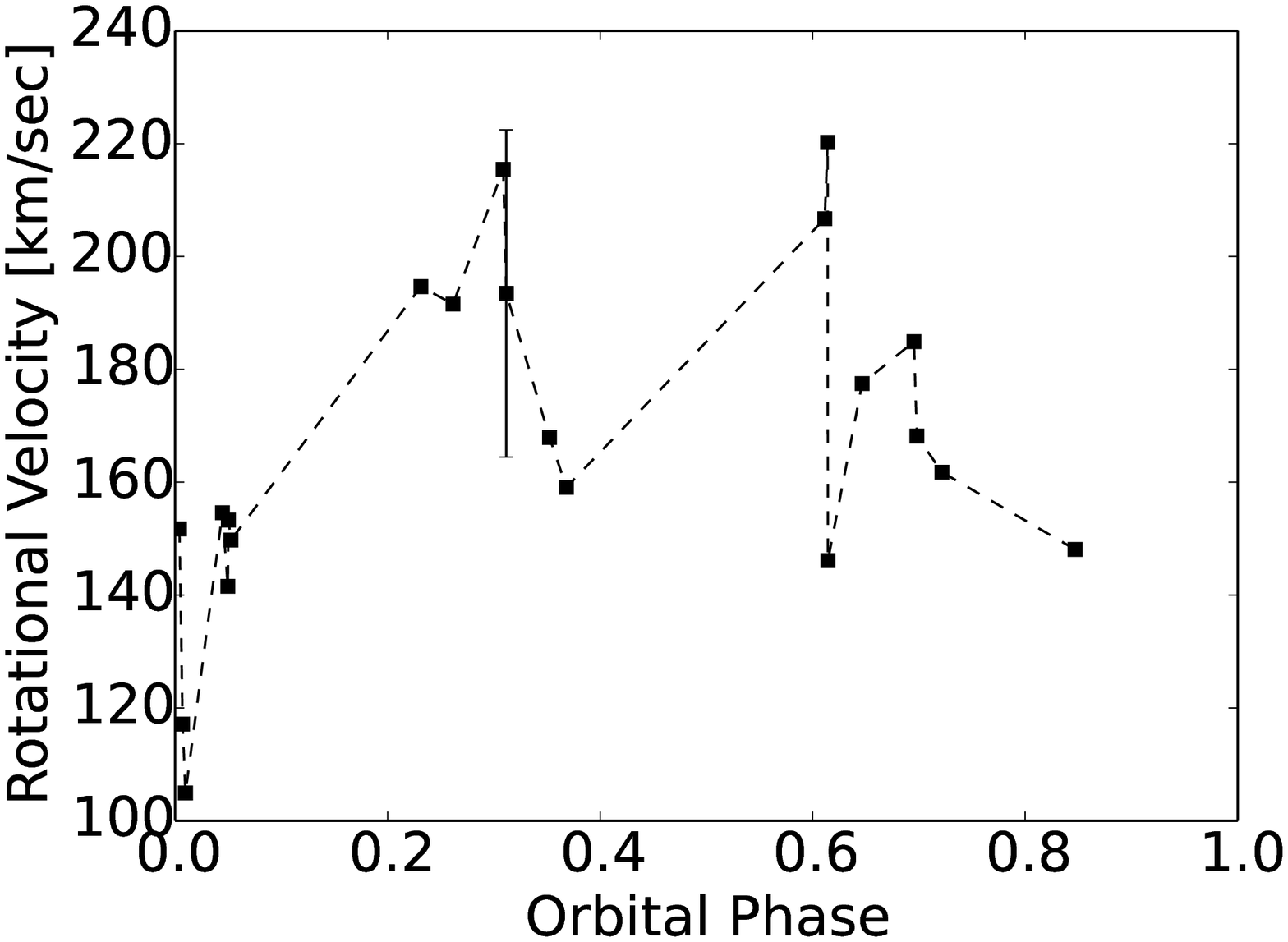} \\
 \end{tabular}$
 \caption{Primary and secondary rotational velocities for $\mu^1$ Sco.
 Due to the Rossiter-McLaughlin Effect the interpretation of values close to phases 0.0 and 0.5 will be uncertain.
}
 \end{center}
 \end{figure}

\subsection{Absolute parameters}
Our derived absolute parameters for $\mu^1$ Sco are listed in Table~12.
\begin{table}
\begin{center}
\caption{Adopted absolute parameters for the $\mu^1$ Sco system. 
Formal errors are shown by parenthesized numbers to the right 
and relate to the least significant digits in the corresponding parameter
values.}
\label{tbl-12}
{\footnotesize
\begin{tabular}{lr}
\hline
\multicolumn{1}{c}{Parameter} & \multicolumn{1}{c}{Value} \\
\hline
Period (days) & 1.44627  \\
Epoch (HJD) & 2449534.177  \\
Mags. $V$, $(B - V)$, $(U - B)$  & 2.98; --0.016; --0.85\\
Vel.\ amplitudes $K_{1,2}$ (km/s)& 140(5); 257(10) \\
Star separation $A$ (R$_{\odot}$)& 12.6(0.5) \\
System vel. $V_{\gamma}$ (km/s)& --6(3) \\ 
Masses $M_{1,2}$ (M$_{\odot}$)& 8.3(1.0); 4.6(1.0)\\
Radii $R_{1,2}$  (R$_{\odot}$) & 3.9(0.2); 4.6(0.3) \\
Surface grav. $g$ (log cgs) & 4.17(0.10); 3.77(0.12) \\
Prim.\ mag. $V_{1}$ & 3.63(0.04) \\ 
Sec.\ mag. $V_{2}$ & 3.85(0.05) \\ 
Prim.\ temp. $T_{e,1}$ (K)& 24000(1000) \\ 
Sec.\ temp. $T_{e,2}$ (K)& 17000(700) \\ 
Distance &  130(20) pc \\
\hline
\end{tabular}
}
\end{center}
\end{table} 

The distance derived from the photometric parallax  (Eqn 3.42 in Budding \& Demircan, 2007), using the data of Table~12, is somewhat
smaller than the HIPPARCOS value of 156 pc (van Leeuwen, 2007), which lies just outside the upper 1$\sigma$ range of the photometric distance estimate.
The photometric estimate has neglected interstellar reddening, however, and, although small at this distance, $A_V$ can be taken to be $\sim$0.07 mag (vAM), which would allow the two distance estimates to agree within the
errors of the determination.  Either distance would be consistent with
the range given for the Scorpius region of the Sco-Cen OB2 Association
(de Zeeuw et al., 1999; Preibisch \& Mamajek, 2008).

 The absolute parameters given in Table 12 are generally 
within their error estimates
of those of vAM, although outside the errors given by vAM.
It can be seen from Figs 12-14 that the scatter in the 
photometric and spectroscopic sets, both vAM's and ours, is $\sim$10\%\ 
that of their amplitudes.  The precision of parameters given by
vAM therefore seem to be over-optimistic,
 as our closer checks confirm -- particularly for the masses.

\section{Discussion}

The  parameters for the stars of GG Lup given in Table~7 suggest
a relatively young pair, and this is supported by the physical characteristics
of the system in which a high rotational velocity of the primary
  has still not been eroded by its frequent close encounters with 
  its companion in the unusually eccentric orbit. 
 In Figure~16 we plot stellar radii versus time using the data of the
 Padova group (Bressan et al, 2012) with a solar composition.  The derived radii and masses
 of the components turn out to be consistent with each other and with an age
 of approximately 33 My. This is not so young as the age estimated by Andersen et al (1993), and indeed significantly 
 greater than representative values for the UCL region
 ($\sim$15 My, Preibisch \& Mamajek, 2008).  
 
 Andersen et al.\ (1993) argued for a metal-composition for the stars of GG Lup significantly below
solar, putting particular weight on their derived relatively high gravity for the secondary star.
Our findings do not give unequivocal support to that argument, even though we have re-analysed 
the photometric and spectroscopic data used by Andersen et al.\ as well as our own data
and find a good measure of agreement in the basic parameters of the fitting functions 
used to match all the data-sets, together with comparable levels of scatter in the residuals.  
Whilst the arguments of Andersen et al.\ are of interest,
particularly in relation to the problem of the origin of Gould's Belt (cf.\ e.g.\ Grenier, 2004),
we believe that a more realistic assessment of the parameter errors,
especially of the masses, do not allow GG Lup, at least not
in isolation,  to be used as a 
reliable discriminant between alternative composition possibilities.
Solar composition models can provide a self-consistent picture for
the observed properties.  In fact, if we were to apply the lower 
metal abundance value considered by Andersen et al.\ ($Z = 0.01$)
the starting radii in Fig 16 are lowered and the
time taken to achieve the presently observed values is increased 
by about 3 My, increasing the disparity with the average age of the UCL region stars.
 But the indications of pulsational instability or anomalously high rotation
  in the envelope of at least the primary component, given the
  unexpectedly persisting orbital eccentricity,
   may render comparisons with normal single stars inappropriate.

\begin{figure}
\label{fig-16}
\noindent\includegraphics[height=8cm,angle=-90]{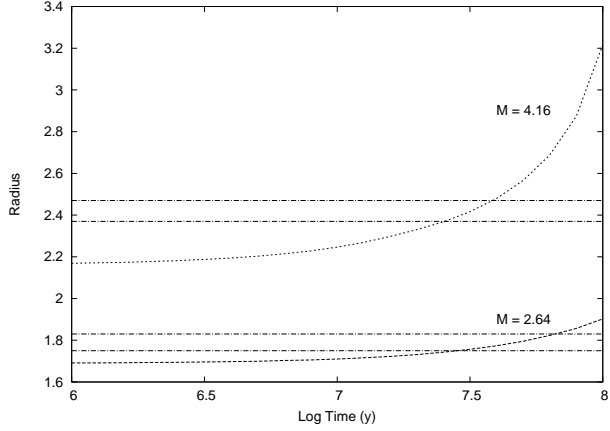}\\
\caption{Radial evolution of the stars in GG Lup.
The 1$\sigma$ error bounds of the measured radii are shown as horizontal lines.}
\end{figure}

\begin{figure}
\label{fig-17}
\noindent\includegraphics[height=8.0cm,angle=-90]{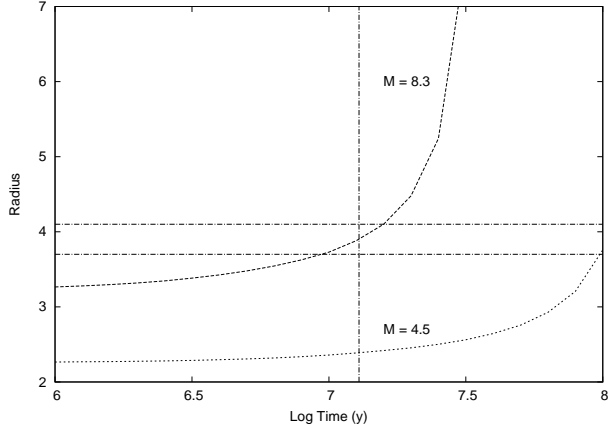}\\
\caption{Radial evolution of a Main Sequence star with the present mass
of the primary of $\mu^1$ Sco.  The 1$\sigma$ limits within its measured
 radius should lie are shown as horizontal lines.  For comparison,
 we show also the radial evolution of a star of mass 4.5 M${\odot}$.}
\end{figure}

On the other hand, for the Sco OB2 Association member 
$\mu^1$ Sco, the radius of a star of the same mass as that found for the primary (8.3 M${\odot}$) agrees closely with that measured
for the star (Fig~17).
This agreement suggests that the primary has been relatively little affected
by the Case A type evolution of the secondary.
The Case A scenario, considered by Giannuzzi (1983), Stickland et al (1996) and others, involving appreciable mass loss from the system in the relatively recent past,
arises because of the anomalous mass, near Roche-lobe-filling, and high-luminosity
secondary star and therefore generally Algol-like condition of the system, yet with the absence of significant period variation.

In that scenario, the secondary must have lost at least
 $\sim$4 M${\odot}$
of its original mass.  The primary's condition, meanwhile, 
remains measurably close to what it would have been as a single star
at its present mass.
If all the lost matter had been transferred, the primary would have started by following the lower track in Figure~17.  But it is difficult to see
how it could have then reached its 
present radius of about 4 R${\odot}$ whilst 
maintaining stability on a Kelvin timescale
in accordance with the insignificant period variation.
A plausible implication is that the very high surface 
temperature of the primary 
causes sufficient ionizing absorption in matter shed from 
the secondary that radiation pressure expels such plasma from the system
via the external Kepler ring (De Loore and Van Rensbergen, 2005).

\section{Acknowledgments}

Generous allocations of time on the 1m McLennnan Telescope and HERCULES spectrograph at the Mt John University Observatory in support of the
Southern Binaries Programme have been made available through its TAC and supported by its  Director, Dr.\ K.\ Pollard and previous Director, Prof.\
J.B.\ Hearnshaw.  Useful help at the telescope were provided by the MJUO management (A.\ Gilmore and P.\ Kilmartin).   Considerable assistance with the use and development
of the {\sc hrsp} software (leading up to the latest version 5) was given by its author Dr.\ J.\ Skuljan. 

Encouragement and support for this programme has been shown by the 
the School of Chemical and Physical Sciences
of the Victoria University of Wellington, as well as the Royal Astronomical
Society of New Zealand and its Variable Stars South section
(http://www.variablestarssouth.org). 
We thank VSS Director Tom Richards for advocating DSLR photometry
in support of the Southern Binaries Programme.

An unnamed reviewer made helpful comments causing us to 
check our results as well as compare them with others more carefully,
resulting in an improved presentation.

{}

\end{document}